\documentclass{raa_rb}
\usepackage{epsfig,lscape,amsmath,amssymb,txfonts}
\usepackage{graphicx,longtable,times,threeparttable,multirow}

\DeclareGraphicsExtensions{.eps,.eps.gz,.epsi}

\begin{document}

\title{The evolution of stellar metallicity gradients of the Milky Way disk from LSS-GAC main sequence turn-off stars: a two-phase disk formation history?}

   \volnopage{ {\bf 2015} Vol.\ {\bf X} No. {\bf XX}, 000--000}
   \setcounter{page}{1}

\author{M. S. Xiang\inst{1},
        X. W. Liu\inst{1,2}, H. B. Yuan\inst{2,4},
        Y. Huang\inst{1}, C. Wang\inst{1}, J. J. Ren\inst{1,4},
        B. Q. Chen\inst{1,4},  N. C. Sun\inst{1}, 
        H. W. Zhang\inst{1}, Z. Y. Huo\inst{3}, A. Rebassa-Mansergas\inst{2,4}
}
   
    \institute{Department of Astronomy, Peking University, Beijing 100871, P. R. China; {\it xms@pku.edu.cn}, {\it x.liu@pku.edu.cn} \\              
     \and
            Kavli Institute for Astronomy and Astrophysics, Peking University, Beijing 100871, P. R. China \\
     \and
              National Astronomical Observatories, Chinese Academy of Sciences, Beijing 100012, P. R. China \\
     \and  
             LAMOST Fellow}

\date{Received~~2015 month day; accepted~~2015~~month day}

\abstract{
          Accurate measurements of stellar metallicity gradients in the radial and vertical direction of the disk and 
          their temporal variations provide important constraints of the formation 
          and evolution of the Milky Way disk. We use 297\,042 main sequence 
          turn-off stars selected from the LAMOST Spectroscopic Survey of the Galactic Anti-center (LSS-GAC) to 
          determine the radial and vertical gradients of stellar metallicity, $\Delta$[Fe/H]/$\Delta$$R$, $\Delta$[Fe/H]/$\Delta$$|Z|$, 
          of the Milky Way disk in the anti-center direction. We determine ages of those turn-off stars by isochrone fitting 
          and measure the temporal variations of metallicity gradients. We have carried out a detailed 
          analysis of the selection effects resulting from the selection, observation and data reduction 
          of LSS-GAC targets and the potential biases of a magnitude limited sample on the determinations of metallicity gradients.  
          Our results show that the gradients, both in the radial and vertical directions, exhibit significant spatial and temporal variations. 
           The radial gradients yielded by stars of oldest ages ($\gtrsim11$\,Gyr) are essentially zero at all heights 
           from the disk midplane, while those given by younger stars are always negative. 
           The vertical gradients deduced from stars of oldest ages ($\gtrsim11$\,Gyr) are negative and show only very weak variations 
           with the Galactocentric distance in the disk plane, $R$, while those yielded by younger 
           stars show strong variations with $R$.  
           After being essentially flat at the earliest epochs of disk formation, the radial gradients steepen as age decreases, reaching 
           a maxima (steepest) at age 7 -- 8\,Gyr, and then they flatten again. 
           Similar temporal trends are also found for the vertical gradients.            
           We infer that the assemblage of the Milky Way disk may have experienced at least two distinct phases. 
           The earlier phase is probably related to a slow, pressure-supported collapse of gas, when the 
           gas settle down to the disk mainly in the vertical direction. In the later phase, there 
           are significant radial flows of gas in the disk, and the rate of gas inflow near 
           the solar neighborhood reaches a maximum around a lookback time of 7 -- 8\,Gyr. The transition of the two phases 
           occurs around a lookback time between 8 and 11\,Gyr. 
           The two phases may be responsible for the formation of the Milky Way thick and thin disks, 
           respectively. And, as a consequence, we recommend that stellar age is a natural, physical 
           criterion to distinguish thin and thick disk stars.             
           From an epoch earlier than 11\,Gyr to one between 8 and 11\,Gyr, there is an abrupt, 
           significant change in magnitude of both the 
           radial and vertical metallicity gradients, suggesting that stellar radial migration  
           is unlikely to play an important role in the formation of thick disk. 
 } 

\authorrunning{M.-S. Xiang, X.-W. Liu, H.-B. Yuan et al.}
\titlerunning{Stellar metallicity gradients of the Galactic disk from LSS-GAC}

\maketitle

\section{Introduction}
\label{sect:intro}

The structure and evolution of the Galactic disk are key issues 
of the formation of the Milky Way, which serves as a unique window 
for the understanding of disk galaxy formation in general. 
With the unique advantage that the individual stars in the Milky Way 
can be resolved, it is possible to address the above issues via a detailed and 
comprehensive archeological approach. 

It has been shown that the stellar density profile in the vertical 
direction of the Galactic disk is best fitted by the sum of two 
exponentials of different scale heights (e.g. Gilmore \& Reid 1983), 
suggesting that the Milky Way disk may contain two components: a thin and a thick disk. 
Similar two-component disks are also observed in external 
galaxies (Burstein 1979; Dalcanton \& Bernstein
2002; Yoachim \& Dalcanton 2008a,b; Elmegreen \& Elmegreen 2006).
Further studies show that the thin and thick disks of the Milky Way are 
different not only in density profiles(Juri\'c et al. 2008; Cheng et al. 2012a), 
but also in kinematics, metallicities and ages. Compared to the thin disk stars, 
the populations of the thick disk are on average older (Fuhrmann 1998; Bensby et al. 2003, 2005; 
Reddy et al. 2006; Haywood et al. 2013), kinematically hotter (e.g. Soubiran et al.  2003), 
and more metal-poor but alpha-enhanced (e.g. Reddy et al. 2003, 2006; 
Bensby et al. 2003, 2005; Fuhrmann 2008).     
On the other hand, evidence has been presented suggesting that the
thin and thick disks may not be two distinct components after all. 
For example, it is shown that the evolution of stellar 
populations resulted from a monotonically declining star formation rate 
can naturally reproduce the observed bimodality of the distribution of $\alpha$-element 
to iron abundance ratios [$\alpha$/Fe] (Sch\"onrich \& Binney 2009a,b). 
Similarly, it has been suggested that the observed 
vertical density distributions of the individual populations selected in the [Fe/H]--[$\alpha$/Fe] 
plane (`mono-abundance populations' -- MAPs)
can be the result of internal evolution of the disk (Bovy et al. 2012). 
The question whether the Milky Way has two distinctive disks, and if so what their 
relation in term of origin remains inconclusive. 

Various scenarios/models have been proposed in recent years to explain 
the formation of the Milky Way disk, especially the thick disks. 
Based on the observed stellar age -- metallicity -- [$\alpha$/Fe] relation, 
Haywood et al. (2013) proposed a two-phase disk formation scenario, 
in which the thick disk forms from well-mixed interstellar medium 
8\,Gyr ago over a time scale of 4 -- 5\,Gyr, and it sets the initial 
chemical conditions for the later, slowly formed inner thin disk. 
Other models proposed to explain the presence of a thick disk invoke    
either external, violent processes or internal, secular evolution. 
In these models  the thick disk results from,  for example, 1$)$ direct accretion of stars in 
infalling satellites 
(Abadi et al. 2003); 2$)$ vertical heating of a pre-existing thin disk 
via minor mergers (Quinn et al. 1993; Villalobos \& Helmi 2008; Read et al. 2008; 
Kazantzidis et al. 2008, 2009; Purcell et al. 2009; Bird et al. 2012); 
3$)$ star formation during a rapid, turbulent disk phase of high gas accretion 
rate at early epoch (Brook et al. 2004, 2005; Bournaud et al. 2009); and 
4$)$ radial migration of stars (Ro$\breve{s}$kar et al. 2008; 
Sch\"onrich \& Binney 2009b; Loebman et al. 2011). 
Those models reproduce well some aspects of the observations, 
yet it remains unclear which model(s) is the most realistic one. 
It is no doubt that the predictions of some of the aforementioned 
models sensitively depend on the input assumptions that are often 
poorly constrained-- for instance, the gas infall/inflow rate and 
the frequency of stellar migration in the radial migration model 
of Sch\"onrich \& Binney (2009b). 
Comprehensive and accurate observations are therefore crucial 
to better constrain those models.  

Over the lifetime of a low mass star, the surface metallicity 
of the star remains largely the same as that at the time of 
formation. Stellar metallicity thus serves as a fossil record of the chemical 
conditions of the Galaxy at the time the star was born. By combining the 
metallicity and age information of a large sample of stars, it is possible 
to reveal the Galaxy formation and evolution history. 
The distribution of the stellar age and metallicity,  
as well as their spatial and temporal variations, 
contain important information of the past history of gas infall, accretion and inflow, 
as well as of the possible merger events and the stellar dynamic evolution 
(e.g. Larson 1976; Quinn et al. 1993; Schonrich \& Binney; 2009a,b). 
This thus serves as a powerful tool to constrain the various scenarios/models 
of Galactic disk formation.

On the one hand, different models/scenarios often predict different 
distributions of stellar age and metallicity, as well as kinematics. For example, 
a thick disk formed via a rapid, gas-rich merger at early epoch   
has a narrow age distribution and almost no vertical metallicity 
gradients (Brook et al. 2004, 2005), 
while a thick disk formed from a slow, pressure-supported collapse, 
following the formation of the extreme Population II stars, 
has an intermediate age distribution and shows negative vertical 
metallicity gradients (Larson 1976; Gilmore et al. 1989).  
A thick disk formed by heating a pre-existing thin disk as a result of minor mergers 
should contain stars older than the age of the current thin-disk (e.g. Quinn et al. 1993), 
while a thick disk arising from stellar radial migration
should contain stars of a wide range of age distribution (e.g. Schonrich \& Binney 2009b).   
On the other hand, many of the aforementioned models are proposed to explain some specific 
aspects of the disk formation, rather than to provide comprehensive predictions 
of the disk properties, including age, metallicity and kinematics. 
It is thus not easy to implement a comprehensive test of those models with observations, 
especially considering the model predictions often sensitively depend 
on the input assumptions. A good knowledge of the stellar age and metallicity 
distributions of a large sample of stars is helpful to set the baseline of the models,   
so as to further tighten their predictions to be tested with observations. 

Metallicity gradients of the Galactic disk have been measured from various tracers, 
including those of relatively young ages, such as OB stars (e.g. Daflon \& Cunha 2004), 
Cepheid variables (e.g. Andrievsky et al. 2002; Luck et al. 2006; Yong et al. 2006),   
H\,{\sc ii} regions (e.g. Balser et al. 2011)
and open clusters (e.g. Friel 1995; Friel et al. 2002; Chen et al. 2003; Magrini et al. 2009), 
and those of intermediate-to-old ages, such as planetary nebulae (PNe; Maciel \& Quireza 1999; 
Costa et al. 2004; Henry et al. 2010), FGK dwarfs  
(e.g. Katz et al. 2011; Cheng et al. 2012b; Boeche et al. 2013), horizontal branch stars 
(Chen et al. 2010, 2011; Bilir et al. 2012) 
and red giants (Hayden et al. 2014). In general, the studies yield negative gradients 
 in both the radial and vertical directions of the disk, along with some fine structures. 
Significant scatters amongst the results, however, do exist. 
Young populations often yield a linear radial gradient of values between $-0.01$ and 
$-0.09$\,dex\,kpc$^{-1}$ (e.g. Lemasle et al. 2013), and there are also evidence
that a linear function is not sufficient to describe the observed radial variations of metallicity 
in the whole disk (e.g. Luck et al. 2003, 2011; Andrievsky et al. 2004; L\'epine et al. 2011).  
From an analysis of the relatively old F turn-off stars targeted in the SEGUE survey (Yanny et al. 2009), 
Cheng et al. (2011b) show that the radial metallicity gradient flattens with increasing 
$|Z|$, the absolute height above the disk plane, and becomes completely flat at $|Z|>1$\,kpc, 
a result that seems to support the rapid, highly turbulent gas-rich merger 
origin of the thick disk at high redshifts (Brook et al. 2004, 2005). 
Similar trend of variations of radial metallicity gradient is also found by Boeche et al. (2013)
from dwarf stars observed by the RAVE survey (Steinmetz et al. 2006) and 
in the Geneva-Copenhagen survey (GCS; Nordst\"orm et al. 2004), as well as by Hayden et al. (2014), 
from red giants targeted in the APOGEE survey (Majewski et al. 2010).  
In the vertical direction of the disk, negative gradients are found with the exact value 
varies from study to study (e.g. $-0.068$\,dex\,kpc$^{-1}$, Katz et al. 2011;
$-0.22$\,dex\,kpc$^{-1}$, Chen et al. 2011; $-0.14$\,dex\,kpc$^{-1}$, Kordopatis et al. 2011; 
 $-0.30$\,dex\,kpc$^{-1}$,  Schlesinger et al. 2012; $-0.243$\,dex\,kpc$^{-1}$,  Schlesinger et al. 2014). 
A strong dependence of the vertical gradient on the Galactocentric 
distance is also found (Hayden et al. 2014).  

To interpret those results, it is helpful to keep several things in 
mind. Firstly, the intrinsic spatial distributions of different tracers 
probably differ significantly. For example, young stars are mostly 
restricted to low Galactic latitudes while older stars are much more spread out. 
Secondly, given that different tracers usually have different characteristic ages, 
they actually probe metallicity gradients of different evolutionary epochs of the Galaxy.  
Thirdly, different studies make use of data from observations 
of different probe limits (in both magnitude and volume) and 
target selection functions. Those effects are responsible, at least partially, 
for the different results obtained in various studies. 
A complete spectroscopic data sample, containing stars that spread over the 
whole age range of the Galaxy, and, at the same time, having a large enough probe volume 
as well as number density for any given age, would be extremely useful 
to avoid/reduce any potential selection effects and draw a clear picture 
of the chemical enrichment history of the Galactic disks.

Obtaining a statistically complete spectroscopic sample of disk stars is, however, 
an extremely difficult task. One obvious obstacle is that the stars are widely distributed  
over the whole sky because of our own location in the Galactic disk, 
thus one needs a survey with extremely large sky coverage to target 
those stars. Secondly, restricted by limited observing capability, only a tiny fraction of the
numerous stars have been spectroscopically targeted hitherto.  
Finally, disk stars often suffer from significant amount of extinction by the interstellar dust grains, 
and, as a consequence, reaching out spectroscopically to a substantial  
distance in the disk is an extremely challenging task. 
Available large sky area spectroscopic surveys are either restricted to the 
solar neighborhood or geared toward halo directions in order to avoid the disk.

The LAMOST Spectroscopic Survey of the Galactic Anti-center (LSS-GAC; Liu et al. 2014) 
aims to collect medium-to-low resolution ($R$\,$\sim$\,1800) optical spectra 
($3700 < \lambda < 9000$\,$\AA$) of more than 3 million stars down to a limiting magnitude of 
17.8\,mag (in $r$-band) in a approximately 3400\,sq.deg. contiguous sky area in the 
Galactic anti-center direction, making use of the newly built LAMOST spectroscopic survey telescope (Cui et al. 2012). 
The survey will deliver accurate stellar parameters (radial velocity $V_{\rm r}$, 
effective temperature $T_{\rm eff}$, surface gravity log\,$g$, metallicity [Fe/H], 
and $\alpha$-element to iron abundance ratio [$\alpha$/Fe] ) derivable from the 
collected  spectra. The contiguous sky coverage, large number of stars and high 
sampling density (100 -- 200 stars per squared degree), as well as the specifically designed 
simple yet nontrivial target selection algorithms  
[random selection in the $(g-r)$ -- $r$ and $(r-i)$ -- $r$ Hess diagrams; Liu et al. 2014; Yuan et al. 2015] 
make the LSS-GAC particularly suitable to study the Galactic disk 
in a star by star yet statistically meaningful manner. 
The LSS-GAC was initiated in September 2012, and is expected to  
complete in 2017. The first LAMOST official data release of the whole LAMOST survey (LAMOST DR1; Luo et al. 2015), 
as well as the first release of value-added catalog of LSS-GAC (Yuan et al. 2015), 
have been recently publicly released. The latter includes stellar parameters yielded by the LAMOST 
Stellar Parameter Pipeline at Peking University (LSP3; Xiang et al. 2015b), 
as well as values of interstellar extinction and stellar distance  estimated with a variety of techniques, 
for a sample of almost another 700\,000 stars targeted by the LSS-GAC by 2013.

In this work, we define a sample of about 300\,000 main sequence turn-off (MSTO) stars 
selected from the first release of value-added catalog of LSS-GAC in an effort to study 
the age-dependent metallicity 
gradients of the Galactic disk. The paper is organized as follows.
In Section\,2,  we briefly introduce the value-add catalog of LSS-GAC. 
In Section\,3,  we introduce the MSTO star sample in detail. 
In Section\,4, we characterize the sample selection effects.
We present our results in Section\,5, followed by a discussion of the implications 
as well as potential uncertainties of our results in Section\,6. 
A summary is presented in Section\,7.

\section{The value-added catalog of LSS-GAC}
Most stars in the current sample are observed during the Pilot- (Oct. 2011 -- Jun. 2012)
and first-two-year (Oct. 2012 -- Jun. 2014) Regular Surveys of LSS-GAC. 
Stellar parameters derived from spectra collected during the Pilot- and first-year 
Regular Surveys have already published 
in the first release of value-added catalog of LSS-GAC (LSS-GAC DR1; Yuan et al. 2015). 
Data from the second-year Regular Survey will be publicly available in the second release of 
value-added catalog of LSS-GAC (LSS-GAC DR2; Xiang et al. in prep.). 

The LSS-GAC DR1 contains stellar parameters derived from spectra of 
signal-to-noise ratios (SNRs) better than 10 for 664\,773 stars, including 
219\,045 stars from the BMF (Bright, Median-bright, Faint) plates toward 
the Galactic Anti-center ($150 < l < 210$\,deg. and $|b|<30$\,deg.), 52\,921 
stars in the M31/M33 area, and 392\,807 stars from the VB (Very Bright;
$r < 14$\,mag) plates (Liu et al. 2014; Yuan et al. 2015). 
The stellar parameters are derived with the LSP3 (Xiang et al. 2015b), and 
have an overall accuracy of 5 -- 10\,km\,s$^{-1}$, 150\,K, 0.25\,dex, 0.15\,dex 
in $V_{\rm r}$, $T_{\rm eff}$, log\,$g$ and [Fe/H], respectively, 
for a blue-arm spectral SNR better than 10 per pixel.
The blue-arm spectral SNR is defined as the median value in the wavelength range 
4600 -- 4700\,{$\AA$}, and one pixel corresponds to 1.07\,{$\AA$} at 4650\,{$\AA$}.
The LSS-GAC DR1 also includes values of extinction and distance 
of individual stars, estimated with a variety of methods. Typical uncertainty is 0.04\,mag  
for the estimates of color-excess $E(B-V)$, 15, 10 and 30 per cent respectively for distances 
of dwarfs, red-clump stars and giants (Yuan et al. 2015). 
 The LSS-GAC value-added catalog for the second-year Regular Survey 
 contains stellar parameters as well as extinction and distance of about another 700\,000 
 stars with a spectral SNR higher than 10.
 
In the current study, we have also included $\sim$\,80\,000 stars 
observed by the Galactic spheroidal parts of the LEGUE survey (Zhao et al. 2012; Deng et al. 2012). 
Those stars have high Galactic latitudes thus suffer little extinction. They were  
originally analyzed to examine the accuracy of the flux calibration pipeline 
of LSS-GAC (Xiang et al. 2015a). The stellar parameters of those stars 
are also measured with the LSP3, and their values of extinction and distance 
are estimated in the same way as for the LSS-GAC targets. Finally, the current 
sample also contains 50\,000 stars in the Kepler field targeted by the LAMOST-Kepler 
Project (De Cat et al. 2015). Those stars are also processed in the same way as for 
LSS-GAC targets.

\section{The main sequence turn-off star sample}
We define a MSTO star sample that contains stars of log\,$g$ around 
the main sequence turn-off. 
The effective temperature of an MSTO star is sensitive to stellar age 
so that it can be used to obtain relatively reliable estimate of the latter using the technique 
of stellar isochrone fitting. 

\subsection{Sample definition}

\begin{figure}
\centering
\includegraphics[width=90mm]{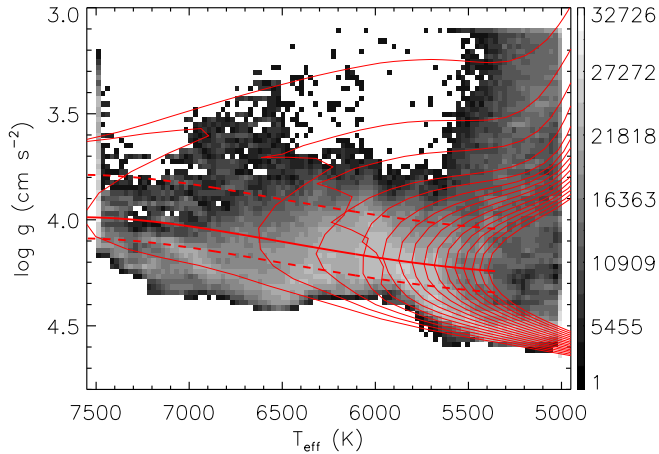}
\caption{Trajectories of MSTO stars of ages, from left to right, 1 to 16\,Gyr, in the $T_{\rm eff}$ -- log\,$g$ 
               plane deduced from the Y$^2$ isochrones of [Fe/H]=0 and [$\alpha$/Fe] = 0. 
                The dashed lines delineate the area of MSTO stars. 
                The background grey-scale image shows the number density of stars of spectral SNR higher than 15 
                 of the current sample.}
\label{Fig1}
\end{figure}

To select MSTO stars, we first define the ranges of values of $T_{\rm eff}$ and
log\,$g$ of MSTO stars of different ages for a given set of [Fe/H] and [$\alpha$/Fe] 
by interpolating stellar isochrones.
As an example, Fig.\,1 shows the trajectories of MSTO 
stars deduced from the Yonsei -- Yale (Y$^2$) isochrones (Demarque et al. 2004) 
in the $T_{\rm eff}$ -- log\,$g$ plane for [Fe/H] = 0 and [$\alpha$/Fe] = 0. 
For a given set of $T_{\rm eff}$ and [Fe/H], we first determine the value of log\,$g$ 
of an {\em exact} MSTO star, log\,$g_{\rm \small TOF}$. Then all stars with the given set of 
$T_{\rm eff}$ and [Fe/H] that have a value of log\,$g$ within the range,  
$a <$ log\,$g$ $-$ log\,$g_{\rm \small TOF}$ $< b$, are selected as MSTO sample stars in the current analysis.
The values of $a$ and $b$ are set to be $-0.2$ and 0.1\,dex, respectively.
A smaller value of $b$ than that of $a$ is adopted in order to reduce the potential 
contamination of dwarf stars.
We further require that spectra of all stars in our sample have a SNR higher than 15, 
and have an effective temperature between 5400 and 7500\,K. The LSP3 yields 
more accurate stellar atmospheric parameters for stars of such SNRs and effective temperature 
range than for stars of lower spectral SNRs or of effective temperatures outside the range 
(Xiang et al. 2015b). In addition, as Fig.\,2 shows, this effective temperature range encloses 
stars of age between 2 and 13\,Gyr for a wide range of metallicities, from 
${\rm [Fe/H]} = -1.0$ to +0.4\,dex. 
A total of 338\,639 unique stars are selected with the above criteria.

\begin{figure}
\centering
\includegraphics[width=90mm]{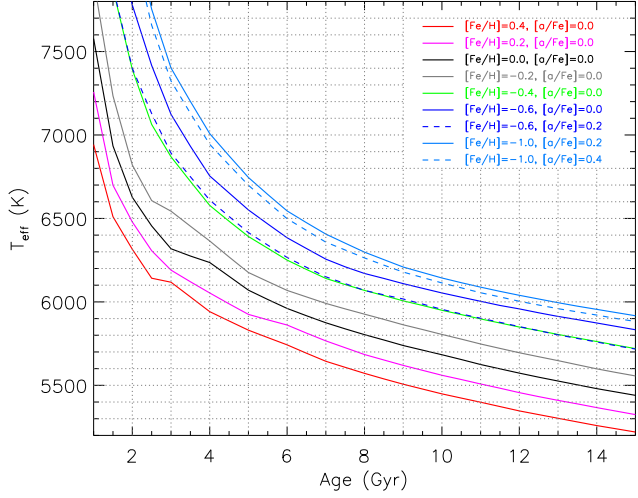}
\caption{Effective temperatures of MSTO stars deduced from the 
                Y$^2$ isochrones as a function of age for different sets of [Fe/H] and [$\alpha$/Fe]. }
\label{Fig2}
\end{figure}

\subsection{The stellar distance and age}
\subsubsection{Determinations of stellar distance and age by isochrone fitting}
We determine distances and ages of MSTO stars by isochrone 
fitting. Given the stellar atmospheric parameters ($T_{\rm eff}$, log\,$g$, 
[Fe/H] and [$\alpha$/Fe]) of a star, the age and absolute magnitude of the star 
can be determined by interpolating stellar isochrones. Since values of 
[$\alpha$/Fe] are not available from the LSP3 yet, we have assumed that our sample 
stars follow the [Fe/H] -- [$\alpha$/Fe] relation inferred for stars in the solar neighborhood 
(e.g. Venn et al. 2004). Stars of ${\rm [Fe/H]}>0.0$ are assumed to have a constant 
[$\alpha$/Fe] of 0.0\,dex; Stars of $-0.5 <{\rm [Fe/H]}< 0.0$ are assumed to have 
an [$\alpha$/Fe] value that increases linearly with decreasing [Fe/H] from 0.0 to 0.2\,dex; 
Stars of $-1.0 <{\rm [Fe/H]}< -0.5$ are assumed to have an [$\alpha$/Fe] value that increases 
linearly with decreasing [Fe/H] from 0.2 to 0.3\,dex, while stars of ${\rm [Fe/H]}<-1.0$
have a constant [$\alpha$/Fe] value of 0.3\,dex.
We have adopted the Y$^2$ isochrones. As a test, the Dartmouth isochrones 
(Dotter et al. 2008) are also used for comparison. Stellar magnitudes of the Y$^2$ isochrones are given 
in filter passbands of the Johnson-Cousin system ($UVBRIJHK$). We transfer them into 
the SDSS and 2MASS systems using the relations presented by Jester et al. (2005) 
and Carpenter (2001), respectively. 

To deduce the distance, the magnitudes are corrected for extinction 
using $E(B-V)$ derived by comparing the measured photometric colors, $g-r$, $r-i$, $J-K_{\rm s}$, 
with synthetic values, as introduced in Yuan et al. (2015), and using the 
extinction coefficients of Yuan et al. (2013). 
For stars in the footprint of XSTPS-GAC, the optical band photometry are 
from the XSTPS-GAC survey (Liu et al. 2014), except for stars 
brighter than 13.0\,mag, for which the data are taken from the
UCAC4 catalog (Zacharias et al. 2013) based on the AAVSO Photometric 
All-Sky Survey (APASS; e.g. Munari et al. 2014), which is  
complete to 15\,mag in $r-$band (Henden \& Munari, 2014). 
For stars outside the footprint of XSTPS-GAC, the optical magnitudes are taken from 
the UCAC4 catalog or the SDSS when available. In our sample, there are a few thousand stars 
that do not have optical photometry from either the XSTPS-GAC, the UCAC4 (APASS) or the  
SDSS surveys. For those stars, only the 2MASS $J, H, K_{\rm s}$ 
magnitudes are used when deriving values of extinction. 
The algorithm used to derive stellar distances is similar to that introduced in Yuan et al. (2015). 
The only difference is that, in Yuan et al., the magnitudes of a star on an isochrone  
that has stellar atmospheric parameters closest to those of the target star are 
simply adopted to estimate the distance of the latter, while in the current work,  we have 
interpolated the isochrones to the desired parameters of our target star of concern. 
The differences in the resultant distances are however marginal in most cases. 

\subsubsection{Calibration and uncertainties of the stellar distances}  
There are several sources of distance errors, including errors of 
the adopted stellar atmospheric parameters, uncertainties of the stellar 
(atmospheric) models that yield the isochrones, and any potential 
mismatches in absolute scale between the LSP3 stellar atmospheric parameters 
and those of the isochrones.  

We estimate the distance errors induced by random errors of the LSP3 
stellar atmospheric parameters by comparing distances deduced from  
multi-epoch observations of duplicate stars in the LSS-GAC sample. 
There are 54\,881 pairs of duplicate observations of spectral SNRs 
better than 10 for which the SNRs of the pair spectra differ by less than 5. 
Those stars are divided into bins of SNR, $T_{\rm eff}$, 
log\,$g$ and [Fe/H], and the mean and dispersion of distance differences 
yielded by the individual pairs of spectra in each bin are calculated. 
The mean values of differences are found to be very close to zero, while 
the dispersions vary with SNR and stellar atmospheric parameters, as one would expect. 
We thus construct a grid of distance errors, taken as the dispersion divided 
by square root of 2, as a function of SNR and stellar atmospheric parameters 
($T_{\rm eff}$, log\,$g$, [Fe/H]).  
For a star of given SNR and stellar atmospheric parameters, the 
distance error resulting from the random errors of LSP3 parameters 
is then estimated by interpolating the grid.  

To quantify systematic errors induced by our method of distance estimation, 
we apply our method to stars in the MILES library (S\'anchez-Bl\'azquez et al. 2006) 
that have accurate Hipparcos distance measurements (Perryman et al. 1997). 
MILES is the template spectrum library used by LSP3 to deduce stellar atmospheric 
parameters (Xiang et al. 2015b). Xiang et al. apply the LSP3 to MILES template 
spectra in order to quantify the systematic errors of the LSP3 algorithms. Here, 
when calculating the isochrone distances of MILES template stars, we have adopted 
the stellar atmospheric parameters  deduced by applying the LSP3 to MILES template 
spectra. By this approach, one can calibrate the resultant distances against the 
Hipparcos measurements to correct for errors introduced by uncertainties of the 
isochrones and by systematic errors of the LSP3 parameters, as well as errors resulting 
from any potential mismatches in absolute scale between the LSP3 and isochrone 
stellar atmospheric parameters. There are 793 MILES stars with distance measurements 
available from the catalog of Extended Hipparcos Compilation (Anderson \& Francis 2012), 
602 of them have distance errors smaller than 10 per cent. 
We divide the 602 stars into bins of $T_{\rm eff}$, log\,$g$ and [Fe/H]. The bin size 
is adjustable to ensure that in each bin there are sufficient number of stars. Normally, 
we require that each bin contains at least 20 stars. An upper 
limit of 600\,K, 0.4\,dex and 1.2\,dex is set for bin size in $T_{\rm eff}$, log\,$g$ and [Fe/H] , respectively. 
As a result, a small number of bins have less than 20 stars. 
The relatively large bin size in [Fe/H] is designed to account for the relatively small 
number of very metal-poor stars. For bins containing more than 5 stars, the mean 
and dispersion of differences between the isochrone
and Hipparcos distances are calculated. Note that for F/G-type stars having 
typical disk star metallicities (e.g. [Fe/H] $< -0.6$), as of interest to this work, there are 
about 20 or more stars in each bin. 
We select bins of $4500 <T_{\rm eff} < 7500$\,K, log\,$g > 3.4$\,dex and ${\rm [Fe/H]}>-1.5$\,dex
that encompass the parameter space of all stars in our MSTO sample, 
and fit the means and dispersions of differences as a function of $T_{\rm eff}$, log\,$g$ 
and [Fe/H] using a third-order polynomial. 
The mean differences as a function of stellar parameters are adopted as 
the systematic errors of our distance estimates, and is corrected for to yield  
our final distance estimates. The dispersions, after combining with the random errors estimated 
above using LAMOST duplicate observations, are adopted as the 
final errors of our distance estimates.

\subsubsection{Calibration and uncertainties of the stellar age}

Errors of age estimates are determined by Monte-Carlo simulations that 
propagate errors of stellar atmospheric parameters as a function of SNR, 
$T_{\rm eff}$, log\,$g$ and [Fe/H], including both random and systematic errors. 

As described in Xiang et al. (2015b), 
the systematic errors actually result from two main causes, one 
is the inadequacy of the LSP3 algorithms -- for instance, the boundary 
effects when estimating log\,$g$. 
The other is the uncertainties in stellar atmospheric parameters of the MILES templates. 
The first cause induces offsets to stellar atmospheric parameters derived with LSP3, and 
the magnitudes of the offsets vary with location in the parameter space, while the second 
cause induces dispersions to the LSP3 stellar atmospheric parameters. 
The systematic errors are estimated by comparing the MILES stellar atmospheric 
parameters and those derived with LSP3 from the MILES template spectra.   
Different to the approach of Xiang et al. (2015b), where the errors induced 
by the above two causes are treated as a whole to derive the final systematic errors,  here 
we determine errors induced by the two causes separately. 
The offsets induced by the first cause are determined in a similar manner 
to that has been carried out for the distance errors as introduced above. 
The LSP3 adopts the weighted mean (for $T_{\rm eff}$, [Fe/H]) or biweight mean (for log\,$g$) 
parameters of the $n$ templates best-matching the target spectrum. Here $n$ 
is larger than 10 for the $T_{\rm eff}$ and [Fe/H] estimation, and $n=8$ for log\,$g$ 
estimation. Thus systematic errors induced by the second cause can be expressed as 
$\sqrt{\sigma^2(X - X_0)/N}$, where $X$ represents the stellar atmospheric parameters 
derived with LSP3 for the MILES templates, $X_0$ represents the MILES stellar atmospheric 
parameters, and $N$ is a number between 1 and $n$. 
Here, we assume that $N=4$. These dispersions are also determined in a similar manner 
to that has been carried out for the distance errors as introduced above. 
The dispersions, combining with the random errors deduced by comparing results from 
duplicate observations as described in Xiang et al. (2015b), 
are adopted to be the error estimates of the stellar atmospheric parameters in the 
current work. The errors are functions of spectral SNR and the stellar atmospheric parameters themselves.

In order to use Monte-Carlo simulations to estimate errors of our age determinations, 
we create a dense grid in the parameter space ($T_{\rm eff}$, log\,$g$ and [Fe/H]).
For each node, random errors are assigned to parameters $T_{\rm eff}$, $\log\,g$ and [Fe/H], 
assuming Gaussian error distributions, with the dispersions of distributions given by 
the above error estimates. With the error-added parameters, 
$T_{\rm eff}$, log\,$g$ and [Fe/H], the age is then determined by isochrone fitting. 
The experiment is repeated for 3000 times for each grid point. 
The deviation of the mean age yielded by the 3000 simulations from the 
desired age that corresponds to the grid parameters $T_{\rm eff}$, $\log\,g$ and [Fe/H], 
combining with offset of age caused by the offsets of LSP3 stellar atmospheric parameters 
as estimated above, is adopted as the systematic error of our age estimate, 
and is corrected for to yield the final age estimates. 
Note that the systematic errors of age arise from two sources, 
one is the systematic errors of stellar atmospheric parameters, and the other 
is the non-linear relation between the age and 
stellar atmospheric parameters and the uneven spacing of the theoretical isochrones. 
The age dispersion (standard deviation) yielded by 
the 3000 simulations is then adopted as the final errors of our age estimates. 
The errors are function of spectral SNR, $T_{\rm eff}$, log\,$g$ and [Fe/H].

A more robust estimation of systematic errors of our age estimates would require a direct 
comparison of our estimated ages with accurate measurements, such as those given by 
asteroseismological analysis. While such an effort is under way, here we concentrate on 
the estimation of relative, rather than absolute values of age.

\begin{figure*}
\centering
\includegraphics[width=140mm]{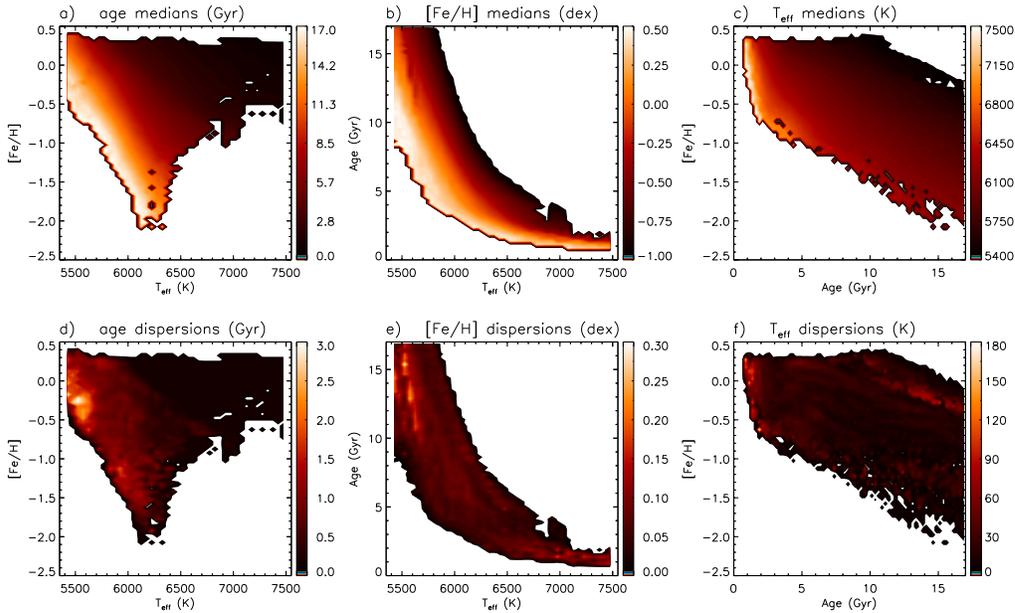}
\caption{Robustness examine for the age estimates. From upper-left to bottom-right, 
               the panels show respectively the distributions of a) age medians in the 
               $T_{\rm eff}$ -- [Fe/H] plane; b) [Fe/H] medians in the $T_{\rm eff}$ -- age plane; 
                c) $T_{\rm eff}$ medians in the age -- [Fe/H] plan; d) age dispersions in the 
                 $T_{\rm eff}$ -- [Fe/H] plane; e) [Fe/H] dispersions in the $T_{\rm eff}$ -- age plane;  
                 and f) $T_{\rm eff}$ dispersions in the age -- [Fe/H] plane.}
\label{Fig3}
\end{figure*}
To check the robustness (consistency) of our age estimates, 
we divide the MSTO sample stars into bins of $T_{\rm eff}$ and 
[Fe/H], and examine the median and dispersion of ages of stars in each 
bin. Similar exercise is also carried out for $T_{\rm eff}$ and [Fe/H], 
respectively.  Fig.\,3 plots the distribution of the resultant 
medians (a) and dispersions (d) of ages in the $T_{\rm eff}$ -- [Fe/H] plane, 
the medians (b) and dispersions (e) of [Fe/H] in the $T_{\rm eff}$ -- age plane, 
as well as the medians (c) and dispersions (f) of $T_{\rm eff}$ in
the age -- [Fe/H] plane. The Figure shows clear trends of variations of 
the median values of each parameter. The dispersions are small in general, 
indicating that our method of age determination is robust. 
If we increase the value of parameter $b$ in Section\,3.1, say from 0.1 to 0.15, 
so as to include more stars in our MSTO sample, we find that while the trends
of variations of the median values remain clearly visible, the dispersions 
become significantly larger at some locations of the parameter space 
as a result of contamination of dwarf stars which have large uncertainties of age estimates.

\subsubsection{Comparison with results deduced using the Dartmouth isochrones}

As a further check of our distance and age estimates, we apply the method using the 
Dartmouth isochrones (Dotter et al. 2008) to our sample stars, and compare the results 
with those deduced using the Y$^2$ isochrones. 
Distances and ages deduced using the two sets of isochrones are compared in Fig.\,4. 
The Figure shows that distances yielded by the two sets of isochrones are consistent 
well with each other. There is negligible systematic differences, with a dispersion 
of only a few per cent. While ages yielded by the Y$^2$ isochrones are 1 -- 2\,Gyr 
systematically younger than those deduced using the Dartmouth isochrones, 
with typical scatters of about 1\,Gyr.  
Results for the age comparison are in agreement with the findings of Haywood et al. (2013).
The systematic differences are probably caused by the different 
stellar physics and/or calibrations adopted in the calculation of 
the two sets of isochrones (Dotter et al. 2008). 
\begin{figure}
\centering
\includegraphics[width=120mm]{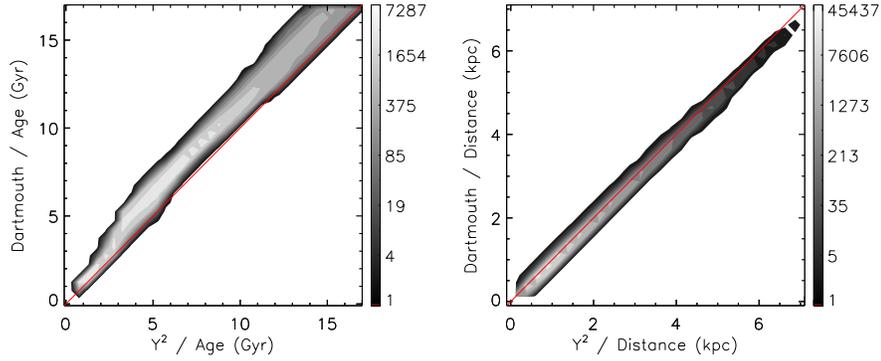}
\caption{Comparison of age (left) and distance (right) yielded by the Y$^2$ isochrones with that 
         by the Dartmouth isochrones for the MSTO sample stars. The grey scale  
         contours show the stellar number densities on a logarithmic scale.} 
\label{Fig4}
\end{figure}

\subsection{Distance, age and spatial position of the sample stars}
\begin{figure*}
\centering
\includegraphics[width=140mm]{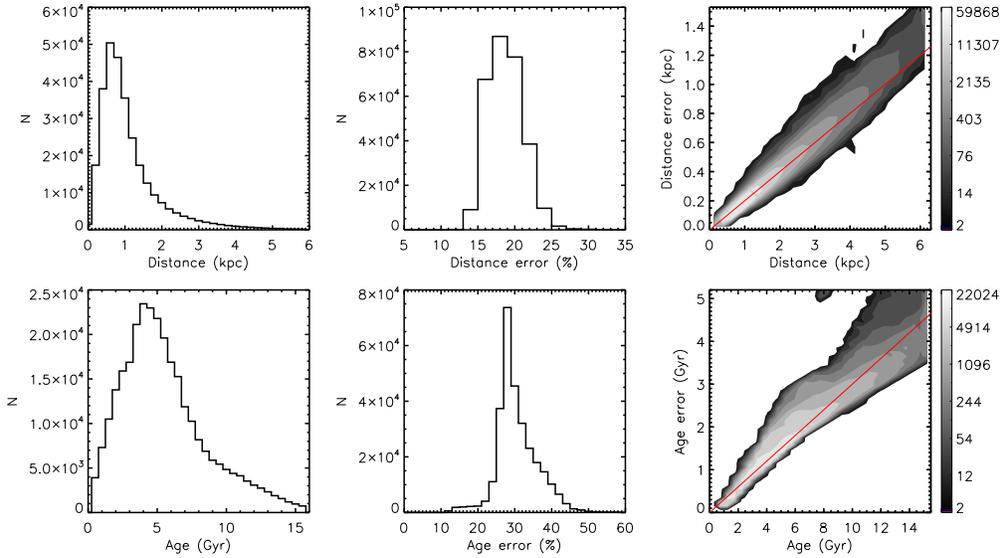}
\caption{Distribution of distances ({\em upper-left}), distance errors ({\em upper-middle}), 
                ages ({\em bottom-left}) and age errors ({\em bottom-middle}) of the MSTO sample  
                stars. The upper-right panel shows the stellar number density distribution in the distance 
                and distance error plane, while the bottom-right panel shows the distribution in 
                the age and age error plane. The grey scale densities are plotted on a logarithmic scale.}
\label{Fig5}
\end{figure*}
Fig.\,5 plots the distributions of distances and ages, as well as their errors, of the MSTO sample stars. 
Most stars are within 2\,kpc. Only about 35\,000 of them fall beyond. Typical distance errors are 20 per cent. 
The ages peak around 4 -- 5\,Gyr, with a tail extending beyond 8\,Gyr. A small fraction of stars 
have ages older than 13\,Gyr or even 15\,Gyr. 
Although isochrones as old as 17\,Gyr are available, stars that old are not 
expected given our knowledge that the universe is younger than 14\,Gyr. 
Thus those old ages found for our sample are likely caused by the errors of stellar atmospheric parameters used. 
We have however opted to keep stars with determined ages as old as 16\,Gyr in our sample 
but discard those older than 16\,Gyr. We believe that most of the rejected, only a small number ($\sim2500$),  
are actually cold dwarfs arising from some problematic log\,$g$ estimates of LSP3. 
Typical age errors are 30 per cent.

Fig.\,6 shows the number density distributions of sample stars in the $R$ -- $Z$ and $X$ -- $Y$ planes. 
Here $X$, $Y$ and $Z$ are coordinates of a right-handed Cartesian coordinate system with 
an origin at the Galactic center. $Z=0$ defines the Galactic disk plane. $X$ points toward the 
Galactic center, $Z$ toward the north Galactic pole. $R$ denotes Galactocentric 
distance in the disk ($X$ -- $Y$) plane, i.e. $R = \sqrt{X^2+Y^2}$.  
The Figure shows that the MSTO sample stars cover well the regions between $7.5 < R < 13.5$\,kpc and 
$-2.5 < Z < 2.5$\,kpc. Stars of high Galactic latitudes are found at larger values of $R$ 
due to less extinction they suffer from.
In the disk plane, most stars fall between $-13.5 < X < -7.5$\,kpc and $-2 < Y < 4$\,kpc.

\begin{figure}
\centering
\includegraphics[width=140mm]{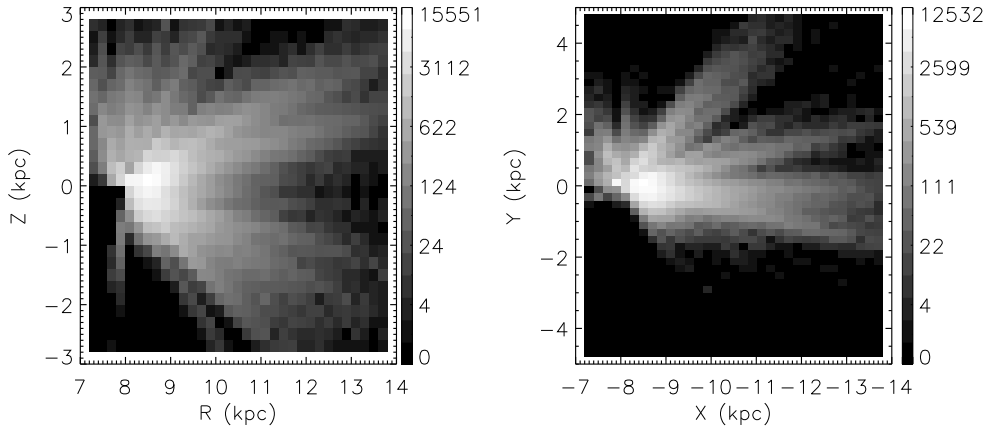}
\caption{Grey scale number density distribution of MSTO sample stars in the $R$ -- $Z$ 
          and $X$ -- $Y$ planes. The stars are binned by 0.2$\times$0.2\,kpc in both the diagrams.
          The densities are shown on a logarithmic scale.}
\label{Fig6}
\end{figure}

\section{Selection effects}

Since not all stars in a given volume are observed, one needs to consider 
the potential selection biases that may be presented in the sample and 
affect our results. 
The current sample has gone through two sets of main selection criteria.
The first is imposed by the magnitude limit of the photometric surveys from 
which the LAMOST targets are selected and get observed. 
The magnitude limits are 9 -- 14, 14 -- 16.5, 16.5 -- 17.8\,mags in $r$-band for the
LAMOST VB, B and M plates, respectively (Liu et al. 2014; Yuan et al. 2015). 
The second criteria are related 
with the SNR as well as $T_{\rm eff}$ and log\,$g$ cuts imposed when selecting 
the MSTO stars from the value-added catalog of LSS-GAC. We have thus implemented a two-steps 
approach to examine the selection effects. First, we calculate the fraction of stars 
selected from the photometric input catalogs that are targeted by the LAMOST and 
have a spectral SNR higher than 15. The calculation is carried out spectrograph 
by spectrograph in the color -- magnitude diagram (CMD) $g-r$ verses $r$, as such 
a diagram is adopted to select the LSS-GAC targets for B, M and F plates. 
Note that the LSS-GAC target selection is in fact based on both $(g-r)$ -- $r$ 
and $(r-i)$ -- $r$ planes (Liu et al. 2014; Yuan et al. 2015), but here we adopt only 
the former to implement the calculation for simplicity. Such a simplification is not expected to 
significantly affect the results given that the stellar locus in the $(g-r)$ and $(r-i)$ 
plane is quite tight (e.g. Ivezi\'c et al. 2007) 
and that we have divided the stars into large color-magnitude cells to calculate the fraction (Section\,4.1).   
Also note that although the selection of VB targets is not carried out  
in the $(g-r)$ -- $r$ plane but based on magnitudes of the targets only, 
our approach is still valid. 
The photometric input catalogs include the XSTPS-GAC, UCAC4 and SDSS 
catalogs. 
Secondly, we examine how the metallicity gradients deduced from the data are 
affected by the selection effects for a magnitude limited sample, utilizing mock data 
generated with Monte-Carlo simulations. 

\subsection{CMD weights}
\begin{figure}
\centering
\includegraphics[width=90mm]{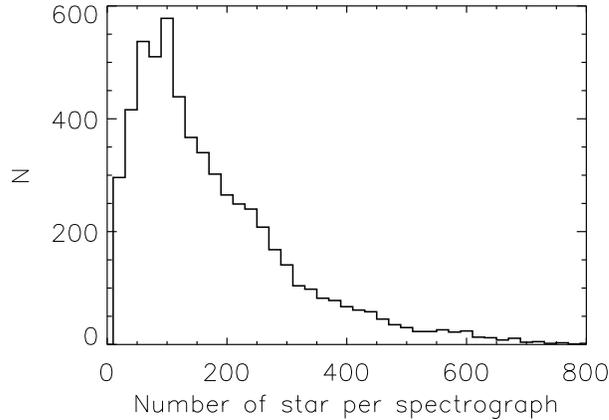}
\caption{Histogram of numbers of stars per spectrograph targeted that have a spectral 
               SNR higher than 15, after combining plates of the same field (see the text).}
\label{Fig7}
\end{figure}

The total sample introduced in Section\,2 contains 1130\,694 unique stars of all colors that have a spectral 
SNR higher than 15. The corresponding spectra are collected from 15\,967 spectrographs of 1126 
plates of 418 fields (Yuan et al. 2015).
For a given field, the projected positions on the sky of the 16 LAMOST spectrographs  
of different plates are nearly the same (differed by a few arc-seconds at most). 
We have thus combined together data from a specific spectrograph from different plates 
for the given field in order to increase the sampling densities. 
Spectrographs containing less than 10 stars are excluded, as well as those 
in which more than 20 per cent of the stars targeted have no counterparts in the 
aforementioned three photometric input catalogs, i.e. the XSTPS-GAC, UCAC4 and SDSS. 
The later cases occur in some few sky areas where the targets are selected from 
other catalogs (e.g. the 2MASS catalog or Kepler input catalog) 
and happen to have no complete photometric data from the XSTPS-GAC, UCAC4 and SDSS surveys. 
After excluding those spectrographs, 1052\,860 stars in 5911 spectrographs remain, 
297\,042 of them are in our MSTO sample.
Fig.\,7 is a histogram of the numbers of stars per spectrograph  which covers 
an FoV of $\sim$1\,sq.deg. The majority (3680) of spectrographs have more than 100 stars. 
The sampling rate (fraction) of a given field depend not only on the total number 
of stars observed in that field but also on the total number of stars in the field within 
the observed magnitude range. 
For fields of high Galactic latitudes in which only VB plates 
are observed, even a relatively small number of stars observed could imply a high completeness 
in the corresponding magnitude range. More than half spectrographs 
that have less than 100 stars are from fields of $|b|>30$\,deg. 

\begin{figure*}
\centering
\includegraphics[width=140mm]{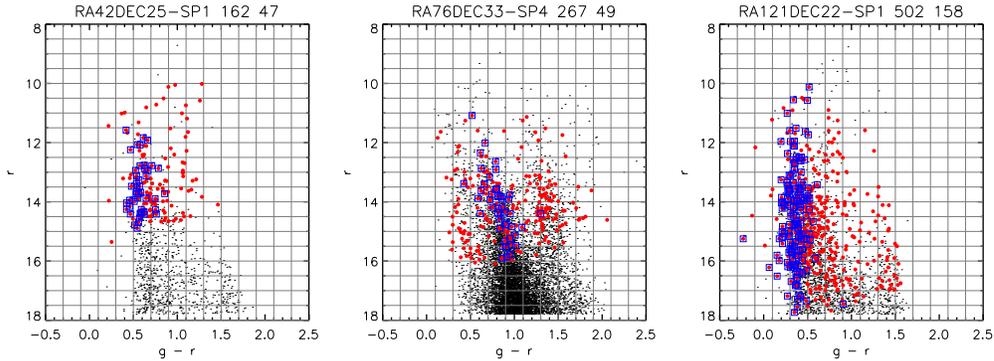}
\caption{$(g-r)$ verses $r$ CMD diagrams for three spectrographs from 
                different fields. Black dots represent point sources in the photometric input catalogs 
                of LSS-GAC, red dots are stars targeted by the LSS-GAC and have a spectral 
                SNR higher than 15, and blue squares are MSTO stars. The field name and spectrograph ID, 
                as well as the number of red dots and blue squares are marked on 
                 top of each panel.}
\label{Fig8}
\end{figure*}
\begin{figure}
\centering
\includegraphics[width=90mm]{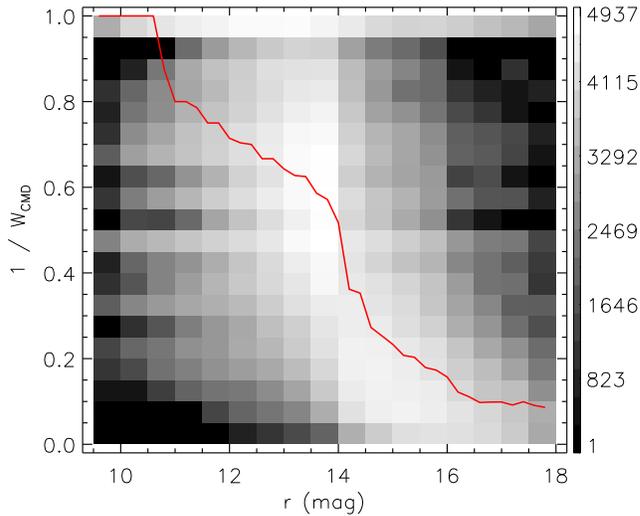}
\caption{ Number density distribution in plane of $r-$band magnitude and the inverse of 
CMD weight for our MSTO sample stars. The grey scale is shown on a logarithmic scale. 
The red line shows the the median value of the inverse of CMD weights of the sample stars 
 as a function of $r-$band magnitude.}
\label{Fig9}
\end{figure}
For each spectrograph, we divide the stars into bins of 0.2$\times$0.5\,mag 
in the $(g-r)$ -- $r$ plane and for each cell calculate the fraction of stars targeted 
by the LSS-GAC with a resultant spectral SNR better than 15 with respect to the 
underlying photometric population. The inverse value of the fraction will be assigned 
as the CMD weights of sample stars in that cell when calculating the mean metallicities 
of the relevant spatial bins where the stars fall inside (see Section\,5). 
This is illustrated in Fig.\,8 for three spectrographs as an example.
To obtain a reliable estimate of the fractions, it is essential that some sufficient 
number of stars are targeted in all CMD grid cells that may contain MSTO stars. 
This is not a problem when the number of stars observed in a given spectrograph 
is large enough, given the fact that the LSS-GAC is designed to target stars of all colors. 
However, it becomes a problem for spectrographs when the number of stars that 
meet the SNR requirement become to small. As a remedy, for spectrographs containing 
less than 50 stars with qualified spectra yet covering the whole magnitude ranges 
of both VB and B plates or B and M plates, the fractions are derived by combining 
data from all spectrographs of  the same field instead of spectrograph by spectrograph. 
Finally, for some spectrographs, some CMD cells near the fainter end of the 
magnitude limit contain only one or two stars with qualified spectra, for example, 
the cell fainter than 16\,mag in the color bin of 0.7 -- 0.9\,mag in the middle panel of Fig.\,8. 
Such stars thus can not be used to represent the average metallicity of stars of the 
photometric sources in the corresponding CMD cells. To treat this problem, for those 
spectrographs, stars belonging to the faintest 2 per cent of all stars in the spectrograph 
of concern are set as outliers. Those stars will be assigned a minimum CMD weight of unity.  
Fig.\,9 plots a grey scale contour map of stellar number density as a function of 
$r-$band magnitude and the inverse of thus derived CMD weight for all the sample stars. 
For the majority of stars, the inverse of CMD weights assigned decrease 
with increasing magnitude. This is due to the fact that in general, the fraction of sample 
stars relative to the underlying photometric population decreases with increasing 
magnitude. More than 90 per cent of the stars have an inverse of CMD weight larger than 0.1 
which corresponds to a CMD weight smaller than 10,  suggesting that the sampling rate 
of our sample is high. The sampling rate is even much higher for the very bright ($r < 14$\,mag) 
stars, for which half of them in the survey fields are included in our sample.

\begin{figure}
\centering
\includegraphics[width=90mm]{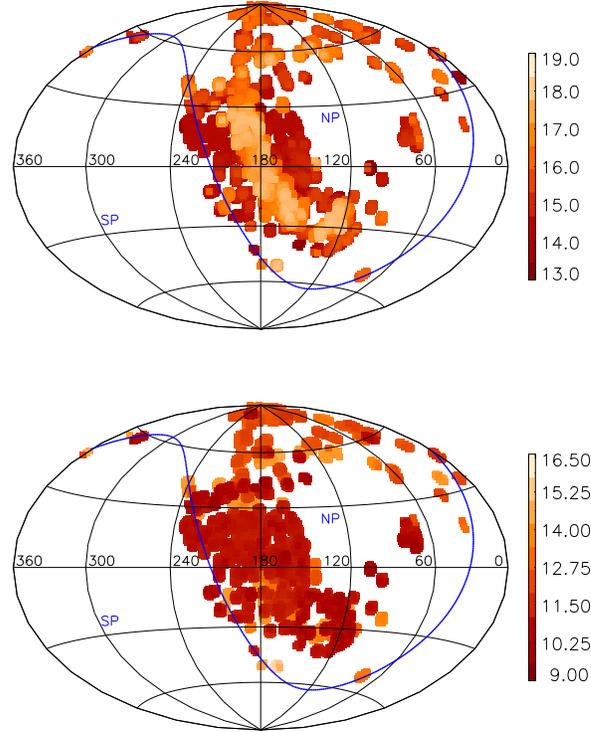}
\caption{Limiting magnitude at the faint (upper) and bright (lower) ends for 
               the individual sight lines (spectrographs).}
\label{Fig10}
\end{figure}
Applying the CMD weights to individual sample stars leads to a magnitude limited sample, 
although due to varying observing conditions and sensitivities of individual spectrographs, 
the limiting magnitudes differ from one sight line to another. 
Fig.\,10 plots the limiting magnitudes at the bright and faint ends 
for the individual sight lines. The limiting magnitudes 
are derived spectrograph by spectrograph, after discarding 
the 2 per cent brightest as well as faintest sample stars in each spectrograph. 
The Figure shows that at the bright end, the majority of sight
lines have a limiting magnitude brighter than 11.0\,mag, while at the faint end, more than half  
sight lines have a limiting magnitude fainter than 16.5\,mag. 
For sight lines for which no VB plates are observed, the 
bright end limiting magnitudes are  fainter than 14.0\,mag.   
  
\subsection{Selection effects of a magnitude limited sample}
In this subsection, we examine potential selection effects that may affect the 
determinations of metallicity gradients using a magnitude limited sample. 
For this purpose, mock data are generated from Monte-Carlo simulations 
of a model disk. The data are then used to investigate how the metallicity 
gradients derived from an observational sample deviate from the assumed 
true values as a result of the limiting magnitudes imposed on the sample. 

Our simple model disk includes two stellar components, a thin and a thick disk. 
The mass density profile of each component is described by a double exponential,  
\begin{equation}
\rho(R, Z) =  \rho(R_{\odot}, 0){\rm exp}(-\frac{(R- R_{\odot}}{R_d}){\rm exp}(-\frac{|Z|+Z_{\odot}}{Z_d}), 
\end{equation}    
where $\rho$ is the mass density, $R_{\odot}$ and $Z_{\odot}$ are the 
position of the Sun in the radial and vertical direction of the disk,  
$R_d$ and $Z_d$ are the scale length and height of the disk, respectively. 
We adopt $\rho(R_{\odot}, 0)$ = 0.05\,${\rm M}_{\odot}$\,pc$^{-3}$, $R_{\odot}$ = 8.0\,kpc 
(Reid 1993), $Z_{\odot}$ = 25\,pc (Juri\'c et al. 2008). The scale parameters 
are from Juric et al. (2008), with $R_d = 2.6$\,kpc and $Z_d=0.3$\,kpc for the 
thin disk, and $R_d = 3.6$\,kpc and $Z_d=0.9$\,kpc for the thick disk. 
For simplicity, we adopt a constant scale length and height 
for stars of different ages.
The mass ratio of the thick disk to the thin disk at $R=8.0$\,kpc 
and $Z=0$\,kpc is adopted to be 0.12 (Juri\'c et al. 2008). 
We adopt the age-metallicity relation as adopted by a recent version of the Besancon 
Galactic model (Robin et al. 2003). In this model, the thick disk forms at 12\,Gyr ago, 
with a mean metallicity of $-0.48$\,dex and a metallicity dispersion of 0.3\,dex, 
while the thin disk forms continuously from 11\,Gyr ago to the current, with the mean  
metallicity increasing from $-0.12$ to 0.01\,dex at the current time and metallicity dispersion 
decreasing from 0.18\,dex to 0.10\,dex at the present epoch. Note that this age-metallicity 
relation is taken from a version in the year of 2011 of the Besancon Galactic model, 
and it is different to the original one presented in Robin et al. (2003). In fact, we find that 
compared to the original one, the age-metallicity relation adopted here is actually 
better consistent with measurement based on our MSTO sample stars in the solar neighbourhood (Xiang et al. 2015, in prep.). 
Metallicity gradients in both the radial and vertical directions are assumed to be zero.       
We assume a constant star formation rate, and at a given age, the star formation 
follows the initial mass function (IMF) of Kroupa (2001). The Y$^{2}$ isochrones 
are adopted to link the stellar physical properties with observables. 
We produce mock catalog with Monte-Carlo simulations, and for each star we 
add the interstellar extinction using the 3-dimensional extinction map of Chen et al. (2014). 
The map covers an area of about 6000\,sq.deg. in the direction of Galactic anti-center. 
For stars outside of those area, 
the extinction map of Schlegel et al. (1998) is used. Finally, for each star, we assign 
a 25 per cent random error to its age, 20 per cent error to distance and 
0.1\,dex error to [Fe/H].    

\begin{figure}
\centering
\includegraphics[width=90mm]{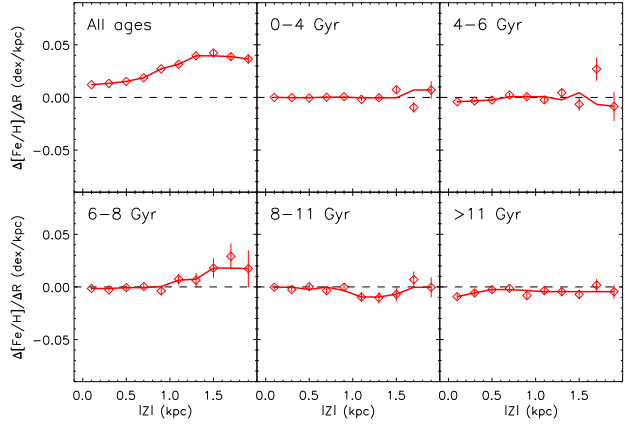}
\caption{Radial metallicity gradients as a function of $|Z|$ derived from the magnitude limited 
         MSTO sample extracted from the mock catalog. The dashed line delineates the gradients (zero) 
         of the model assumptions.  
         The individual panels show results derived from stars of different
         age bins as marked in the plots.} 
\label{Fig11}
\end{figure}

\begin{figure}
\centering
\includegraphics[width=90mm]{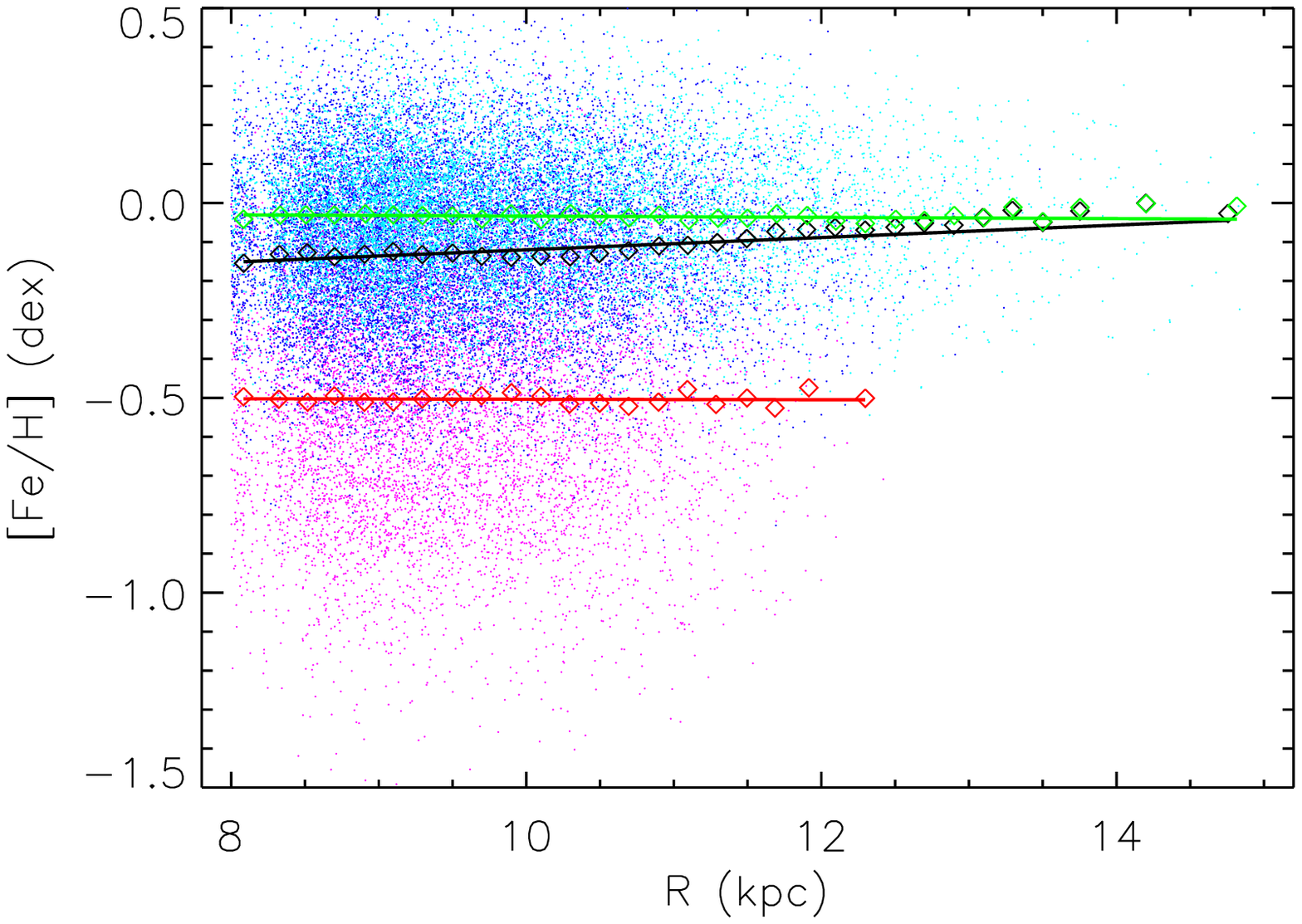}
\caption{The plot illustrates the selection effects that lead to false, non-zero radial metallicity gradients
         deduced from a magnitude limited sample of stars of $|Z| < 0.1$\,kpc. The data are drawn from the  
          mock catalog. Dots of different colors represent stars of different ages, with cyan ones being the 
          youngest and red ones the oldest. 
         Diamonds in green and red are respectively the mean metallicities  
         of stars having an age of 2\,Gyr and 12\,Gyr at different Galactic radii, while black 
         diamonds represent the mean metallicities of stars of all ages. The error bars 
         associated with the mean metallicities are tiny and therefore not plotted. 
          The lines are linear fits to the data points.
          Because of the limiting magnitudes imposed on the sample stars, there are few old, metal-poor stars 
          at large distances.}
\label{Fig12}
\end{figure}

We extract MSTO stars from the mock catalog that fall inside the FoVs and limiting 
magnitudes as our magnitude limited sample derived in the previous subsection 
in the anti-center direction ($R>8.0$\,kpc), and compare the metallicity gradients 
derived from the mock sample with the model assumptions (i.e. zero gradients). 
Fig.\,11 plots the radial metallicity gradients thus derived from the mock sample 
as a function of absolute height above the Galactic plane ($|Z|$) for stars of different ages. 
Details of the definition and derivation of radial metallicity gradients are introduced in Section\,5.1. 
The Figure shows that gradients derived from stars of all ages deviate from the model 
input which is set at zero, marked by dashed lines in the plot. 
The deviations are artifacts caused by the different volumes probed by different 
stellar populations (ages and metallicities) for the given limiting magnitudes. 
As an example, this is illustrated in Fig.\,12. Younger, more metal-rich stars 
have higher turn-off effective temperatures, thus higher (intrinsic) luminosities 
than older, more metal-poor stars. Given the limiting magnitudes 
imposed on the sample stars, younger, more metal-rich stars reach larger 
distances than those older, metal-poor ones, leading to an artificial 
positive metallicity gradient. The magnitude of this false gradient  
depends on the metallicity distribution of stars in the volume 
studied. As the height ($|Z|$) increases, the fraction of metal-poor 
(thick disk) stars increases, yielding larger deviations. At $|Z| > 1.5$\,kpc, 
metal-rich (thin disk) stars are so few such that they play unimportant role 
in the derivation of metallicity gradients. 
As a result, the deviation from the true value of the gradients derived actually 
decreases for heights above 2\,kpc. 
    
Fortunately, for sample stars in a narrow age bin, such artifacts are found 
to be almost absent. Fig.\,11 shows that metallicity gradients derived from stars 
of the individual age bins reproduce well the underlying assumption for $|Z| < 1.5$\,kpc. 
Here the age bins are defined in the same way as those for the observed MSTO star sample 
(Section\,5.2). Note that in deriving the gradients, stars of [Fe/H] $< -1.0$\,dex have been discarded 
as for the case of the observed sample. The cut is to avoid contamination from halo stars (Section\,5.1). 
At $|Z| < 1.5$\,kpc, the largest (absolute) gradients derived from the mock data 
are about 0.01\,dex\,kpc$^{-1}$. The deviations from the true, zero gradient start to 
occur near the outer edges of height probed by the 
corresponding stellar populations. Near the edges, the selection biases 
induced by the metallicity effects, i.e, for a given age, more metal-rich 
stars are more luminous and are thus abundant in the observed sample, 
and vice versa, become important. At a given age, a 0.5\,dex difference 
in metallicity may lead to about 0.2\,mag difference 
in absolute magnitude and about 10 per cent difference in distance, 
comparable to the bin size at $|Z|\sim$1.5 -- 2\,kpc.   
The selection effects are complicated given the fact that our sample have different 
limiting magnitudes for different sight lines. 
In our simulation, the maximum deviations of radial gradients are about 0.02 -- 0.03\,dex\,kpc$^{-1}$,  
found for heights $1.5<|Z|<2$\,kpc from stars in the 6--8\,Gyr age bin (Fig.\,11). 
Such deviations are unlikely to pose a major problem for our analysis given 
their small magnitudes and the fact that they occur only near the edges of 
volumes probed by the sample stars.
Note that for our sample, bias induced by the bright end limiting magnitudes 
is negligible given that most FoVs of the sample contain very 
bright stars ($r\sim10$\,mag). The maximum deviation caused by the 
bright end magnitude cut is about 0.01\,dex\,kpc$^{-1}$, as shown the first data point 
in the last panel ($>11$\,Gyr) of Fig.\,11.

\section{Results}
\subsection{Spatial distribution of mean metallicities}
\begin{figure*}
\centering
\includegraphics[width=140mm]{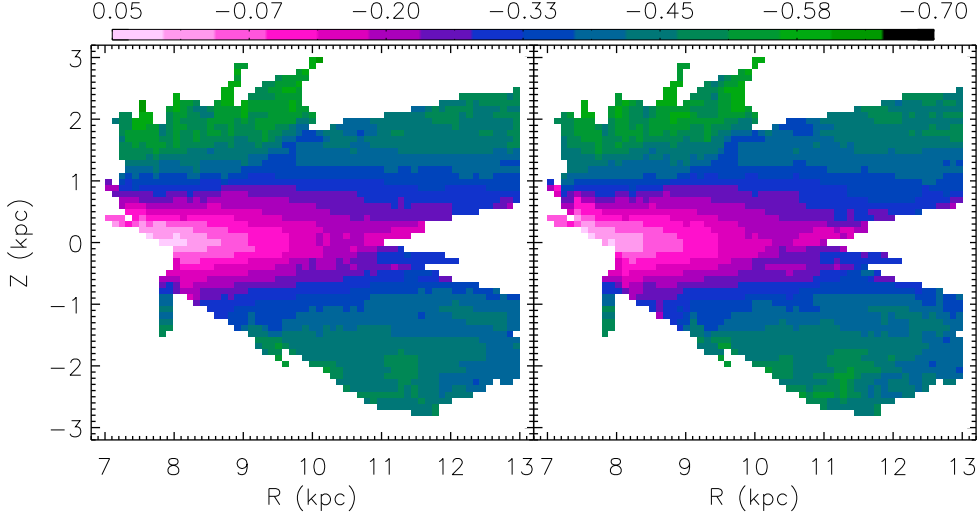}
\caption{Distributions of mean metallicities in the $R$ -- $Z$ plane before (left) and after (right) applying CMD weights for our sample stars.}
\label{Fig13}
\end{figure*}
We divide the sample MSTO stars into bins in the $R$ -- $Z$ plane. 
The bins step by 0.2 and 0.1\,kpc in the $R$ and $Z$ directions, respectively, 
with bin sizes determined by 0.2+$|R-R_\odot|\times$0.1 
and 0.2+$|Z|\times$0.1\,kpc in the $R$ and $Z$ directions, respectively, to account 
for distance errors which become larger at larger distances. 
For each bin, we calculate the mean metallicity, defined as, 
\begin{equation}
{\rm [Fe/H]}_{\rm mean} \equiv \int_{\rm [Fe/H]_1}^{\rm [Fe/H]_2}{n({\rm [Fe/H]})\times{\rm [Fe/H]}d{\rm [Fe/H]}}
\end{equation} 
where $n({\rm [Fe/H]})$ is the normalized, fractional stellar number density as a function of [Fe/H], 
[Fe/H]$_1$ is set to $-1.0$\,dex to avoid contamination of halo stars, 
and [Fe/H]$_2$ is set to infinity. Note that the current sample contains only 4000 stars 
of ${\rm [Fe/H]} < -1.0$\,dex, and no stars of ${\rm [Fe/H]} > 0.5$\,dex. 
We use summation in replacement of the integration in actual calculation.  

Fig.\,13 shows a color-coded distribution of the mean metallicities in the $R$ -- $Z$ plane. 
The left and right panels show respectively the results derived before and after applying the 
CMD weights deduced in Section\,4.1.
When plotting the distributions, we have discarded bins containing less than 30 stars. 
The distributions before and after CMD weight corrections are quite similar, suggesting 
that the CMD corrections have only a marginal effects on the mean metallicities. 
A strong negative radial metallicity gradient is clearly seen near the disk 
plane, and the gradient flattens as one moves away from the midplane. 
There are also significant negative vertical gradients, which 
become shallower as one moves outward. The distributions are largely  
symmetric with respect to the disk plane. 
To quantitatively characterize the spatial and temporal variations of 
mean metallicities, we derive the radial and vertical metallicity 
gradients and study their temporal variations in the remaining part of this Section.

\subsection{Radial metallicity gradient}

\subsubsection{The radial gradients as a function of $|Z|$}
\begin{figure*}
\centering
\includegraphics[width=140mm]{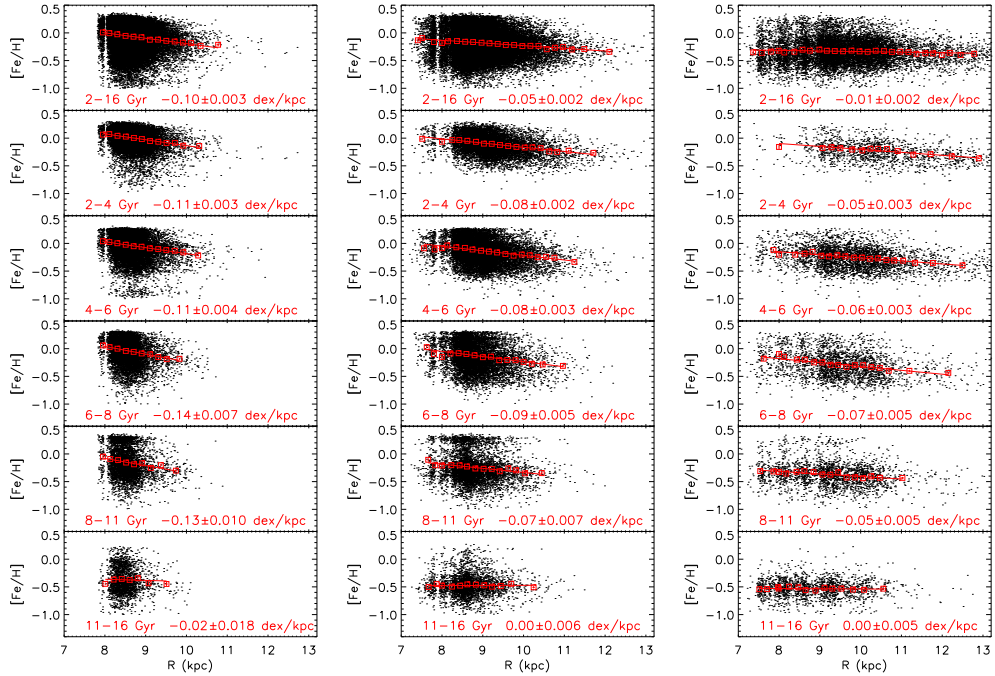}
\caption{Radial metallicity gradients $\Delta$[Fe/H]/$\Delta R$ derived from stars of 
          $|Z|<0.1$\,kpc (left panels), $0.4<|Z|<0.6$\,kpc (middle panels) and $0.9<|Z|<1.1$\,kpc (right panles)  
          and of different age bins, as marked in each panel. Red squares represent mean metallicities in the individual  
           radial bins, after applying the CMD weights. The error bars of the mean metallicities are 
           overplotted, but in all cases they are smaller than the size of the symbols given the large number 
           of stars in each radial bin. 
           Red lines are linear fits to the red squares.  The 
           fitted gradient and associated error thus derived are marked in each panel.  }
\label{Fig14}
\end{figure*}
\begin{figure}
\centering
\includegraphics[width=80mm]{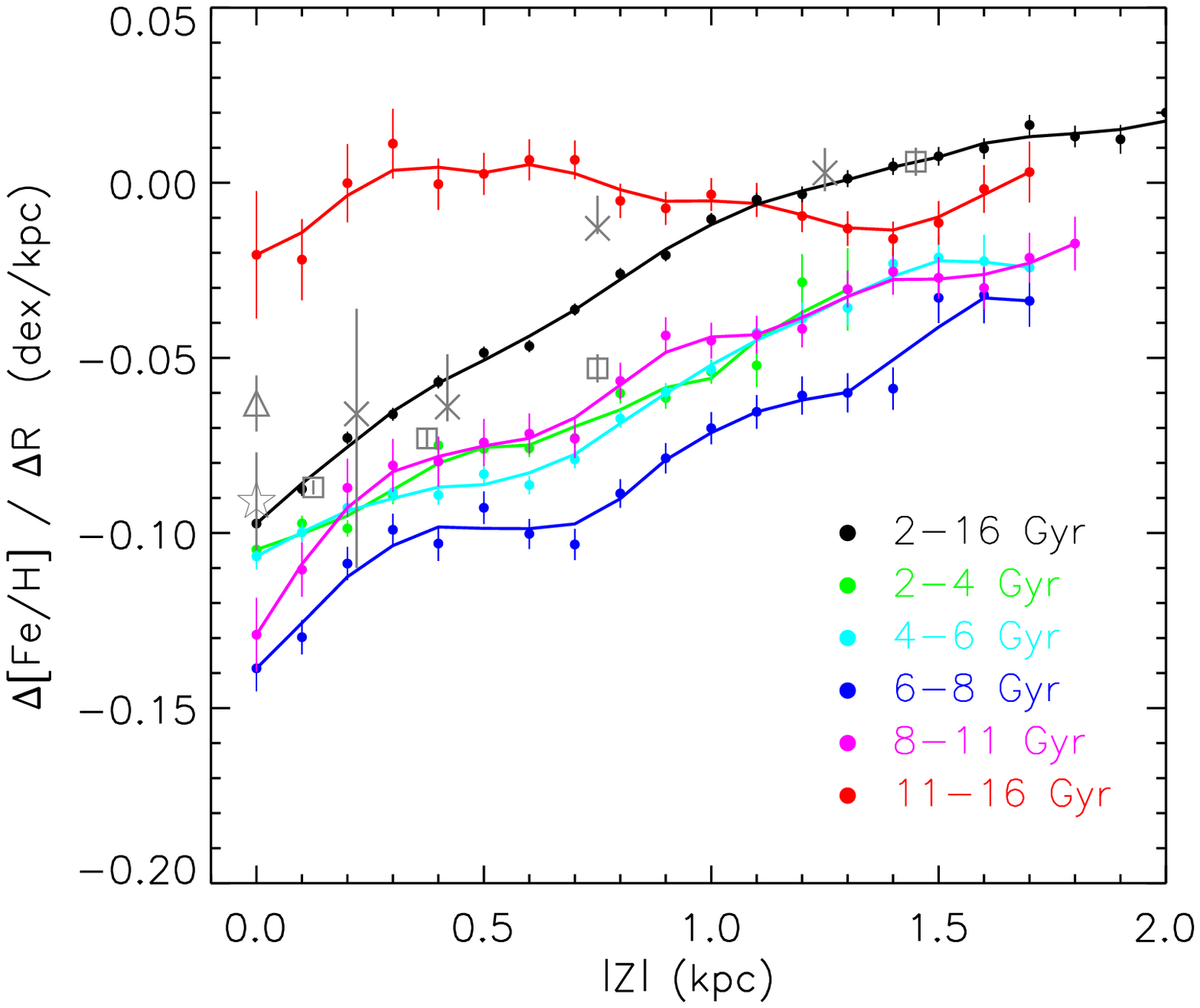}
\caption{Radial metallicity gradients as a function of $|Z|$
          derived from stars of different age bins. Dots represent actual measurements, 
           while lines are drawn by smoothing across 3 adjacent dots. 
           Different colors represent results from stars of different age bins, as 
           marked in the plot. Grey symbols are measurements from the literature. 
           The triangle represents the measurement of Chen et al. (2003) using open clusters as tracers,  the 
           star is the measurement of Friel (1995), again using open clusters, the 
           crosses are measurements of Cheng et al. (2012b) derived from SEGUE turn-off stars, 
           and the squares are from Hayden et al. (2014) deduced from APOGEE giants.}
\label{Fig15}
\end{figure}
We divide the stars into slices of $|Z|$ of thickness of 0.2\,kpc. In each slice, the mean 
metallicities of the individual bins in radial direction as a function of $R$ are fitted with a linear function. 
The slope of the fit is adopted as the radial metallicity gradient.
In each slice, we divide the stars into radial bins 
by requiring that each radial bin contains a minimum of 100 stars. 
The minimum bin size is set to be 0.2\,kpc.
For stars in each radial bin, we calculate the mean metallicity and 
uncertainty, after applying the CMD weights to the individual stars, 
as well as the error-weighted mean and error of $R$. A linear function 
is then used to fit the mean metallicities as a function of $R$ to derive the radial 
metallicity gradient. Errors of the mean metallicities and 
the mean values of $R$ are taken into account in the fitting to estimate the 
error of the derived radial metallicity gradient. 

In doing so, we divide the MSTO sample stars into groups of 
different age bins. Considering that the metallicity of the very young stars are 
probably biased against metal-poor ones due to an upper limit cut in $T_{\rm eff}$ 
at 7500\,K (cf. Fig.\,2), we discard stars younger than 2\,Gyr to avoid the potential bias. 
The remaining stars are binned into groups of ages: 2 -- 4\,Gyr, 4 -- 6\,Gyr, 
6 -- 8\,Gyr, 8 -- 11\,Gyr and 11--16\,Gyr. 
For each age group, we derive the radial metallicity gradients in  
different $|Z|$ slices. Fig.\,14 shows the results for three $|Z|$ slices, 
$|Z| <0.1$\,kpc, $0.4 < |Z| < 0.6$\,kpc 
and $0.9 < |Z| < 1.1$\,kpc. The Figure shows that within a given $|Z|$ slice, 
stars of older ages are more metal-poor and exhibit shallower radial metallicity gradients.   

The derived radial metallicity gradients as a function of $|Z|$ are plotted 
in Fig.\,15. The Figure shows that the radial gradients have significant spatial and temporal variations. 
The gradients derived from stars of all ages (2--16\,Gyr) increase monotonously, 
from $-0.1$\,dex\,kpc$^{-1}$ at $|Z| = 0$\,kpc to $0.02$\,dex\,kpc$^{-1}$ at $|Z| = 2$\,kpc. 
However, stars of different age bins present significantly different gradients. 
In particular, stars of the oldest ages show essentially zero gradients at $all$ heights, 
while stars of younger ages have negative gradients and the gradients from 
individual age bins are always steeper than those derived from 
stars of all ages. Except for those of the oldest age bin, gradients derived from 
stars of the individual age bins show significant spatial and moderate temporal 
variations: the gradients flatten with increasing height above the plane, and, 
as the lookback time decreases, the gradients first steepen, reaching a maximum 
(negative) value around 6 -- 8\,Gyr, and then become shallower again. 
The gradients derived from stars of some age bins (e.g. those of 6--8\,Gyr and 8--11\,Gyr) 
 also show some fine features as $|Z|$ varies. 
 The genuineness of those fine features is hard to assess for the moment given the potential 
selection effects of the current sample which has limiting magnitudes that vary from one 
sight line to another (cf. Section\,4.2).  

\subsubsection{The radial gradients as a function of age}
\begin{figure*}
\centering
\includegraphics[width=140mm]{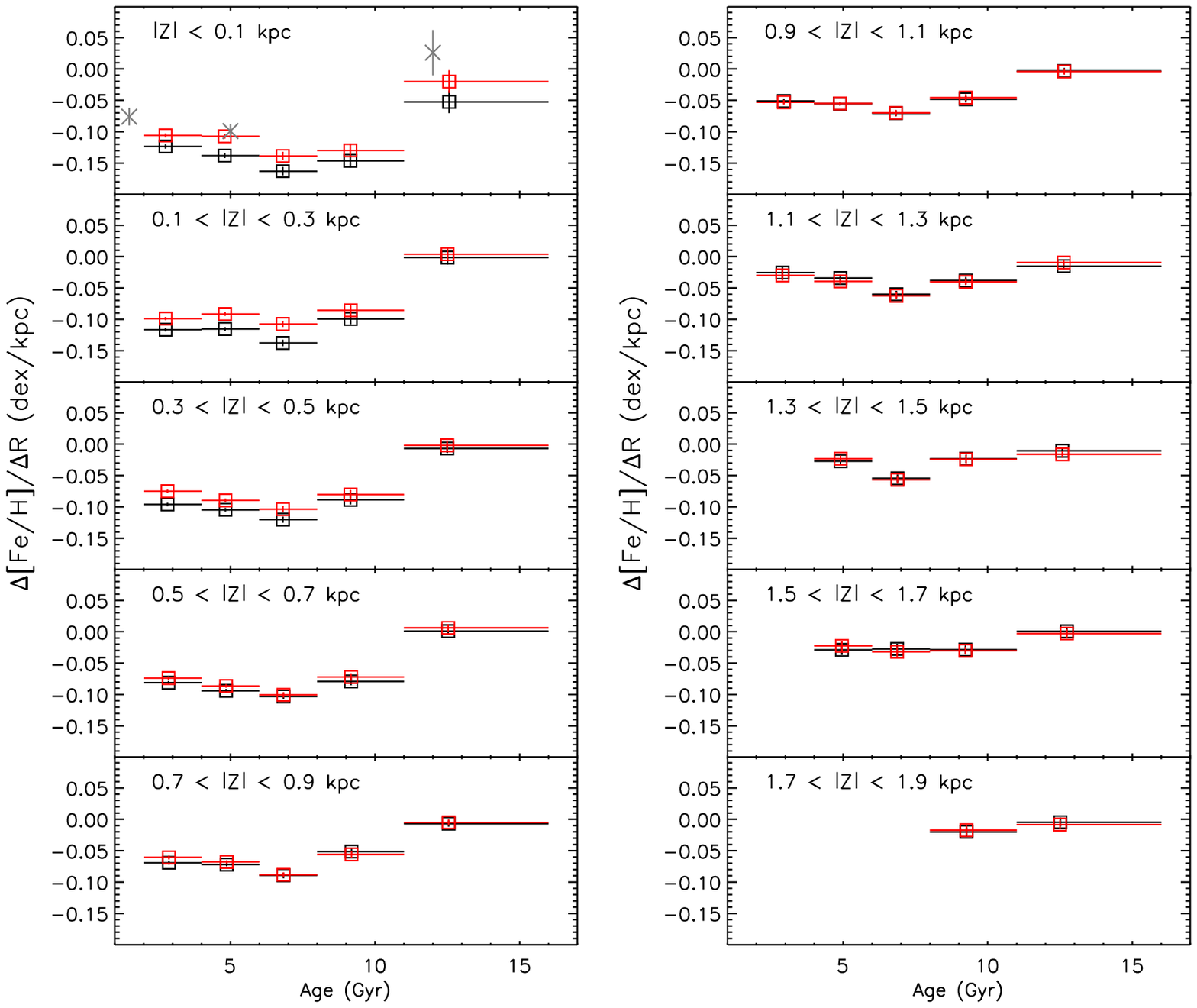}
\caption{Radial metallicity gradients as a function of age for stars
         in different $|Z|$ bins, as marked in the plot. The black and red squares
         represent respectively results before and after applying the CMD weights to our sample stars. 
         The horizontal error bars represent the age ranges of stars adopted to derive 
         the radial gradients.
         The vertical error bars are the fitting errors of radial gradients, and 
         in most cases they are very small because of the large number of stars. 
         They grey crosses in the $|Z| < 0.1$\,kpc bin are measurements 
         from Nordstr\"om et al. (2004).}
\label{Fig16}
\end{figure*}

Fig.\,16 plots the radial metallicity gradients as a function of age for 
stars in different $|Z|$ slices. It shows that at the earliest epochs (age $> 8$\,Gyr), 
the gradients steepen with decreasing age, reaching a maximum around 6 -- 8\,Gyr, 
then flatten again. Similar trends are seen in all height slices of $|Z| < 1.5$\,kpc. 
As already shown by Fig.\,15, the negative gradients 
of stars younger than 11\,Gyr flatten as $|Z|$ increases, while the 
gradients of the oldest ($> 11$\,Gyr) stars are essentially $zero$ and invariant with $|Z|$. 
As a result,  stars at large $|Z|$ show only a weak trend of variations with age at the early epochs. 
For the height slice of $0.9 < |Z| < 1.1$\,kpc, the gradient steepens from 
a value of 0.0\,dex\,kpc$^{-1}$ around 12.5\,Gyr to $-0.07$\,dex\,kpc$^{-1}$ at 7\,Gyr, corresponding an 
evolution rate of $-0.013$\,dex\,kpc$^{-1}$\,Gyr$^{-1}$. The corresponding rate is $-0.020$\,dex\,kpc$^{-1}$\,Gyr$^{-1}$
for height slice $0.3 < |Z| < 0.5$\,kpc and $-0.022$\,dex\,kpc$^{-1}$\,Gyr$^{-1}$ for $|Z|<0.1$\,kpc. 
Analyses show that the radial gradients derived from stars 
of 7 -- 9\,Gyr are comparable with those from stars of 6 -- 8\,Gyr, 
indicating that the gradients probably reach the maximum earlier, probably 8\,Gyr ago. 
This suggests that the evolution rate of the radial gradient at early epochs 
of the disk formation is even faster. Unfortunately,  given the relatively large  
uncertainties and poor (absolute) calibration of our current age determinations, 
it is difficult to pin down the exact time that the radial 
gradients reach the maximum and the exact values of the 
evolution rate of the gradients.  For stars of ages younger than 8\,Gyr, 
as the age decreases, the gradients flatten and the evolution rate slows down to 
about 0.007\,dex\,kpc$^{-1}$\,Gyr$^{-1}$, much slower than the rates at  
the early epochs for bins of low $|Z|$.    
The application of CMD weights to the data mainly changes the gradients in slices of lower $|Z|$ ($<0.5$\,kpc), 
and leads to an increase of gradients between 0.01 and 0.03\,dex\,kpc$^{-1}$. 
Nevertheless, the CMD correction does not change the overall trend of gradients as a function of age.

\subsubsection{Comparison with previous results}
In Fig.\,15 we have also plotted measurements of the radial gradients from 
several recent studies, including the measurement of Cheng et al. (2012b) 
using  SEGUE F turn-off stars, that of Hayden et al. (2014) deduced from APOGEE 
red giants, and those of Friel (1995) and Chen et al. (2003) using open clusters as tracers. 
The Figure shows that the gradients and their variations with $|Z|$ given by 
those earlier studies are largely consistent with our results deduced from 
stars of all ages, except for the bin of $|Z| = 0.75$\,kpc where the measurement 
of Cheng et al. is about 0.02\,dex\,kpc$^{-1}$ higher (less steep) while that of 
Hayden et al. is lower by a similar amount (more steep). 
At $|Z|\sim$0.1\,kpc, the gradient,  $-0.063\pm0.008$\,dex\,kpc$^{-1}$, derived from an analysis 
of 118 open clusters by Chen et al. (2003), seems to be too high (less steep), 
by as much as 0.03 dex\,kpc$^{-1}$, compared to our result as well as that of Friel (1995), 
who find a gradient of $-0.091\pm 0.014$\,dex\,kpc$^{-1}$ using 44 open clusters between 7 and 16\,kpc. 

We show in Section\,4 that metallicity gradients derived from stars of a wide range 
of ages suffer from strong selection effects. 
In general the effects lead to shallower (negative) metallicity gradient 
than the true value (cf. Section\,4.2), or even a positive gradient at large $|Z|$.  
Both our gradients derived from stars of all ages 
and those of Cheng et al. are likely suffered from such effects. 
On the other hand, as we show in Section\,4, our gradients deduced from stars of 
the individual age bins should be hardly affected by those effects.  
Hayden et al. (2014) divide their sample into stars of high-[$\alpha$/Fe], 
which they believe to be mainly consist of old stars, and stars of low-[$\alpha$/Fe], 
which they believe to be mostly young stars. For the young stellar population, 
they find a radial metallicity gradient of $-0.03\pm0.006$\,dex\,kpc$^{-1}$ 
for stars of heights $1.0 < |Z| < 2.0$\,kpc. This value is basically in agreement with 
our results for stars younger than 11\,Gyr in the same height range.
For the old, high-[$\alpha$/Fe] population, Hayden et al. find that the radial metallicity 
gradients show significant variations with $|Z|$, in contrast to our results. For stars 
of the oldest ages ($> 11$\,Gyr), we find that the radial gradients are essentially zero 
at all heights covered by the current sample ($<2$\,kpc). 
Most open clusters have ages younger than 2\,Gyr, and they generally yield a 
negative radial gradient ranging, for example, from $-0.09$\,dex\,kpc$^{-1}$ 
(Friel \& Janes 1993; Friel 1995) to $-0.06$ dex\,kpc$^{-1}$ (Friel et al. 2002; Chen et al. 2003).  
Parts of the scatter of results may be caused by the different ranges of $R$ and $Z$, 
maybe also of age covered by the different samples of open clusters employed in the individual studies. 
As described earlier (Section\,5.2.1), to avoid bias, we have excluded stars younger than 2\,Gyr 
in our sample. Our results deduced from stars older than 2\,Gyr show that the gradients 
flatten as age decreasing in the past few Gyr. Thus the relatively shallower gradients yielded by 
open clusters compared to our measurements from young stars near the disk plane  
is probably not surprising. Note also that the sample employed by 
Chen et al. (2003) includes open clusters spreading over a wide range of age and $|Z|$. 
This may have also caused the relatively shallower radial gradient found by them.    

In Fig.\,16 we have also plotted the measurements of Nordstr\"om et al. (2004) 
 in the panel of $|Z|<0.1$\,kpc (top-left), considering that their stars are limited 
to a small volume of the solar neighborhood. 
Note that Nordstr\"om et al. use the mean orbital radii rather than the measured 
values of $R$ in deriving the gradients. 
The overall trend of variations of gradients with age deduced from this very local sample 
of stars is in good agreement with that reported here, although there are some differences 
in their absolute values.
Nordstr\"om et al. (2004) suggest that the gradients of young disk 
stars steepen mildly with age, but stars of the oldest ages ($> 10$\,Gyr) do not 
follow the trend. In fact, results from PNe also suggest that the radial metallicity 
gradients of the Galactic disk steepen with age (Maciel et al., 2003; Maciel \& Costa, 2009). 
All these are consistent with what we see in the current data, in much more detail.

\subsection{Vertical metallicity gradients}
\subsubsection{Vertical gradients as a function of $R$ and age}
\begin{figure}
\centering
\includegraphics[width=80mm]{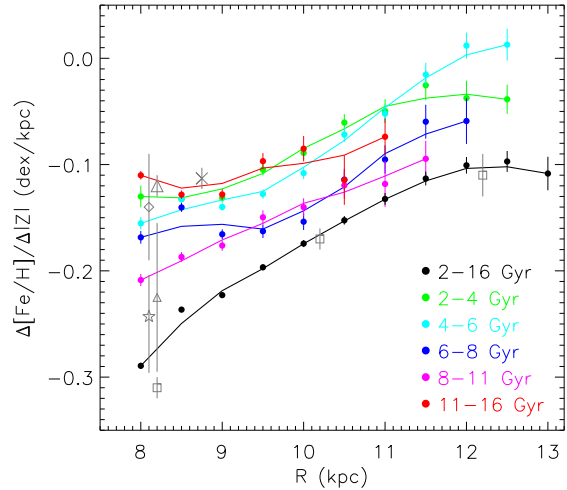}
\caption{Same as Fig.\,15 but for vertical gradients as a function of $R$.
         The grey symbols are measurements from APOGEE giants (squares) by 
         Hayden et al. (2014), SEGUE stars (five pointed star) by Schlesinger et al. (2014), 
         SEGUE FGK dwarfs of $1<|Z|<3$\,kpc (cross) by Carrell et al. (2012),  
         thick disk stars of $1<|Z|<4$\,kpc (diamond) by Kordopatis et al. (2011), 
          SEGUE RHB stars of $|Z|<3$\,kpc (small triangle), and from 
          thick disk RHB stars of $0.5<|Z|<3$\,kpc (large triangle) by Chen et al. (2011).}
\label{Fig17}
\end{figure}

To derive the metallicity gradients along the vertical direction, we divide 
the stars into annuli of 1\,kpc in the radial direction. In each annulus, 
we further divide the stars into bins of $|Z|$ by requiring that each 
vertical bin contains at least 100 stars that cover at least 0.2\,kpc 
in the vertical direction. The slope of a linear fit to the CMD weight corrected mean 
metallicities as a function of $|Z|$ is adopted as the vertical gradient 
for that radial annulus. As in the case for radial gradients, we divide the 
sample stars into groups of different age bins to examine the temporal 
variations of vertical gradients.   

Fig.\,17 plots the vertical gradients derived from stars of different 
age bins as a function of the mean radius $R$ of the radial annulus. 
It shows that for stars of almost all age bins, the vertical gradients flatten 
as $R$ increases. The only exception are those derived 
from the oldest ($> 11$\,Gyr) stars -- the vertical gradients yielded by those old stars 
are largely constant, at the level of about $-0.11$\,dex\,kpc$^{-1}$, becoming 
only slightly shallower by about 0.02\,dex\,kpc$^{-1}$ from 8 to 11\,kpc.
In fact, considering the uncertainties of the gradients and the systematics induced by 
potential unaccounted for selection effects
the vertical gradients derived from the oldest stars are 
essentially consistent with no radial evolution at all. Phenomenologically, it is probably 
not surprising that one sees such trends of variations of vertical 
gradients as a function of $R$ given that we have already shown that, 
except for those of oldest ages, our sample stars exhibit negative values of 
radial metallicity gradients that become shallower as $|Z|$ increases, 
while for those oldest stars the radial gradients are essentially zero at all heights (cf. Figs.\,15, 16).

Fig.\,17 also shows that the vertical gradients derived from stars of all ages
are significantly steeper than those derived from stars of the individual age bins.
This result reflects the differences of the spatial distribution of stars of different ages. 
Young stars, mostly metal-rich, concentrate at small heights, while those of
older ages, often more metal-poor stars, are more often found at larger heights. 
This leads to a steeper negative vertical gradients if stars of all ages are put together
to derive the gradients. Note that this is different from the case for the 
radial gradients, where if one includes stars of all ages when determining the 
metallicity gradients, then the selection effects will lead to shallower gradients 
than the true underlying values.

At a given $R$, the vertical gradients show significant evolution with ages. 
The oldest stars have a nearly constant vertical gradient of about $-0.11\pm0.02$\,dex\,kpc$^{-1}$.
The gradients become much steeper at age bins 8 -- 11\,Gyr and 6 -- 8\,Gyr,  
and then become shallower again. This is seen in all annuli of $R<11$\,kpc. 
In the radial annulus of $R = 9.0$\,kpc,  where the sample 
suffers from the least selection bias as a result of the magnitude cuts both at the faint and 
the bright ends, it seems that 8\,Gyr is an apparent turning point of the vertical gradients. 
It is difficult to assess at the moment whether the epoch of this turning point 
of vertical gradients varies with $R$ or not due to the potential 
systematics of our current measurements at both small and large radii. 
However, the presence of a turning epoch around 8\,Gyr, with the exact time 
that may differ by one or two Gyr, seems to be secure.
At $R = 9.0$\,kpc, the vertical gradients steepen at a rate of about 0.01\,dex\,kpc$^{-1}$\,Gyr$^{-1}$ 
from the earliest epoch ($\sim$12.5\,Gyr) to 8\,Gyr, and then flatten with a comparable speed 
at later epochs. 

\subsubsection{Comparison with previous results}

In Fig.\,17, we also show some recent measurements of the vertical gradients, 
including those of Chen et al. (2011) utilizing SEGUE red horizontal branch (RHB) stars, 
Kordopatis et al. (2011) utilizing thick disk stars in the solar neighborhood,  
Carrell et al. (2012) and Schlesinger et al. (2014) utilizing SEGUE FGK stars, 
as well as Hayden et al. (2014) utilizing APOGEE red giants. 
Note that since in many of those studies the exact  
mean radial position of sample stars in not given, we have inferred a rough value 
based on the published plots.
The comparison shows that the values of vertical gradients derived from the current sample of MSTO stars 
of all ages are well consistent with determinations of Hayden et al. (2014), who find a 
value of about $-0.31$, $-0.17$ and $-0.11$\,dex\,kpc$^{-1}$ at $R = 8$, 10 and 
12\,kpc, respectively. From a sample of RHB stars of $|Z| < 3.0$\,kpc, 
Chen et al. (2011) find a vertical gradient of $-0.225$\,dex\,kpc$^{-1}$, comparable to 
that found by Schlesinger et al. (2014) and consistent with our result found 
from stars of all ages if one takes into account  the relative large uncertainties of 
measurement of Chen et al. Note that after excluding the thin disk and halo stars, 
Chen et al. find a vertical gradient of $-0.12$\,dex\,kpc$^{-1}$ from the remaining 
thick disk stars of $0.5 < |Z| < 3$\,kpc. 
This result is corroborated by Schlesinger et al. based on a similar exercise. 
Similarly, Carrell et al. (2012) find a vertical gradient of $-0.113$\,dex\,kpc$^{-1}$ 
from SEGUE FGK dwarfs of $1 < |Z| < 3$\,kpc, where the 
thick disk stars are expected to dominate. 
Here for stars of the oldest ages, we find a gradient of $-0.12$\,dex\,kpc$^{-1}$.
All these results seem to support our conclusion that the vertical gradients 
show significant temporal variations. They also support
our conjecture that those stars of oldest ages in our sample are actually thick disk stars (cf. Section\,6).

\section{Discussion and implication}

In this section, we discuss the uncertainties of our results, 
as well as the implication of our findings of metallicity gradients 
on the nature of the thick disk and how that may constrain the formation 
and evolution history of the thin and thick disks. 

\subsection{Uncertainties of measurements}
There are two main error sources that introduce uncertainties to our determinations
of metallicity gradients, the errors of stellar atmospheric parameters and the
potential selection effects hidden in a magnitude limited sample.

Errors in the stellar atmospheric parameters lead to uncertainties
in the estimated distances and ages, and introduce contamination, say by dwarfs, in our
MSTO star sample. The largest uncertainties are probably come from log\,$g$ estimates.
The LSP3 values of log\,$g$ adopted in the current work suffers from the boundary effects (Xiang et al. 2015b).
Some moderate systematic errors are also found to exist in the adopted log\,$g$
via a comparison with the asteroseismic values (Ren et al.  2015, in prep.).
These systematic errors have not been corrected for in the current work. 
To account for the potential systematics in distance estimates induced 
by errors in the stellar atmospheric parameters, we have calibrated the 
distances derived using the Hipparcos distance measurements of the MILES 
spectral template stars. The solution works well for MSTO stars, which constitute
the majority of MILES stars in the temperature range $5400 < T_{\rm eff} < 7500$\,K.
However, there are some contaminations in our MSTO star sample from dwarf or subgiant
stars as a result of the boundary effects for log\,$g$ determinations.
For those contaminating stars, their distances are either over- or underestimated,
and their ages are always overestimated. Such effects mainly occur near the
low temperature cut, where stellar isochrones of different ages are closely
packed in the $T_{\rm eff}$ -- log\,$g$ diagram (Fig.\,1).
Near the low temperature end ($T_{\rm eff} < 5500$\,K), there are also some weak
clumping of metal-rich stars in the $T_{\rm eff}$ -- [Fe/H] plane (e.g. Fig.\,42 of Yuan et al. 2015a),
which are possibly artifacts in the LSP3 parameters, caused by the clustering and holes
in the distributions of MILES template stars in the parameter space. The clumping of metal-rich stars
seen in the 8 -- 11\,Gyr age bins of Fig.\,14 are in fact contributed by those stars.
A comparison with the LAMOST DR1 parameters show that those stars are also
metal-rich ([Fe/H] $> 0$\,dex) in the LAMOST DR1, though their distribution is
smoother in the space of LAMOST DR1 parameters than in that of LSS-GAC parameters.

To further investigate that how our results may have been affected by distance errors, 
we reassign distances to the MSTO sample stars based on their distance errors using 
a Monte Carlo technique and redo the analysis presented in Section\,5. 
We have performed 500 simulations. In most cases, the resultant gradients 
as well as the trends of their spatial and temporal variations show no significant changes 
compared with those presented in Section\,5, suggesting that our results are not significantly 
affected by the distance errors. The only noticeable change is that 
some of the fine features seen in the trends of radial gradients as a function $|Z|$ (Fig.\,15) 
disappear in some of the simulations, suggesting that those fine features are probably 
artifacts caused, at least in parts, by the distance uncertainties.

We have implemented several tests to examine the potential effects of contamination
induced by dwarfs and subgiants in our sample. First, we continuously
shrink the $T_{\rm eff}$ range of our MSTO star sample to reduce the level of contamination,
and then redo the analysis. We find that even with a very restrict range of 
$5700 < T_{\rm eff} < 6500$\,K, the data still yield results basically consistent with 
those presented in the current work. This implies that contamination from dwarfs and 
subgiants probably does not have a big impact on our study. 
There, however, are some noticeable changes in the results of radial gradients 
at $|Z|<0.4$\,kpc deduced from stars older than 6\,Gyr, as they become flatter. 
Those changes are understandable. With a $T_{\rm eff}$ cut of 5700\,K, 
the sample become significantly biased in the sense that old, metal-rich stars 
have been systematically excluded from the sample, leading to 
an artificially flatter radial gradients. Nevertheless, the changes do not 
affect our main conclusion that the metallicity gradients show significant spatial 
and temporal variations, and that the radial metallicity gradients of 
the oldest stars are essentially zero at all heights.
Our other tests include repeating the current
analysis using a MSTO star sample selected from the official LAMOST DR2, as well as
using new LSP3 parameters estimated with a KPCA method, which will be implemented
in the next release of LSP3 parameters, to be included in the value-added catalog of LSS-GAC DR2.
We find that measurements of stellar atmospheric parameters yielded by different pipelines
give results consistent with what we presented in the current work.
We thus believe that uncertainties in stellar atmospheric parameters adopted in the current work
do not affect our results significantly. More accurate parameters will however be very useful to
characterize the temporal variations of metallicity gradients, both in radial and vertical directions,
in particular to pin down the exact turning epoch when the evolution of gradients
changes from a mode of steeping with decreasing age to a mode of flattening with decreasing age,
as well as the exact epochs when the thick disk formation started and ceased.


As discussed in Section\,4, our sample is a magnitude limited one.
The sample is biased against metal-poor stars near the farther distance limit
corresponding to magnitude cut at the faint end and biased in favour of
more metal-poor stars near the nearer distance end corresponding to the
magnitude cut at the bright end. The magnitudes of those effects vary from
one sight line to another. The effects may be responsible for some of
the fine features seen in Figs.\,15--17.
We expect that as the survey progresses, the sky coverage of the VB, B and M plates
continues to improve, such effects will be much less important.

\subsection{Thin and thick disks -- two different phases of the disk formation?}

We show that sample stars of the oldest ages ($>$\,11\,Gyr) have 
essentially zero radial metallicity gradients at all heights, completely 
different to the behaviour of stars of younger ages. The latter exhibit  
negative radial gradients that flatten with increasing height. In the direction 
perpendicular to the disk, stars of oldest ages show vertical gradients 
almost invariant with $R$, while stars of younger ages have gradients 
that flatten significantly with increasing $R$. 
These results strongly suggest that the stars of the oldest ages have experienced 
a formation process very different to stars of younger ages. 
It seems that those oldest disk stars
formed from gas that is well-mixed in the radial direction, while younger stars 
formed from radially inadequately mixed gas. It is widely accepted that gas 
accretion and inflow play a key role in providing gas and determining  
the star formation history of the Galactic disks. Gas inflow is expected 
to transfer metals inwards, increasing the radial (negative) metallicity 
gradients of the disk (e.g. Lacey \& Fall 1985; Sch\"onrich \& Binney 2009a). This suggests that gas accretion 
and inflow that determine the star formation and enrichment history of the younger disk 
did not play a significant role in the formation of the old disk. 
Rather the old disk seems to have been formed via some violent, fast processes, 
such as gas collapse or merging.    

The existence of a thick disk (Gilmore \& Reid 1983) of the Milky Way was first 
proposed more than 30 years. Over the years, much have been learned 
with regard to the properties of the thick disk and various criteria have been proposed 
to characterize the thick disk stars, such as the spatial position in the 
vertical direction $|Z|$, the 3-dimensional velocity with respect to the local standard 
of rest (LSR) and the metallicity [Fe/H] and $\alpha$-element to iron abundance ratio [$\alpha$/Fe]. 
To distinguish thick disk stars from thin disk stars based on their spatial positions alone is difficult. 
At low Galactic latitudes, the disk is dominated by thin disk stars. Even at relatively high 
latitudes, the contaminations of thin disk stars could be substantial. The kinematics of 
disk stars are affected by a variety of processes (churning and blurring) related with asymmetric 
(sub)structures (the bar, spirals, and stellar streams). Samples of thick disk stars selected 
purely based on stellar kinematics are also often biased in metallicity since there seems to be 
a correlation between the velocity and metallicity of a disk star due to secular evolution of the 
Galactic disk (e.g. Lee et al. 2011). 
As for metallicity, there seems to be no physical base to give a clear cut that distinguish 
the thin and thick disk stars. In the current, we find that stars of the oldest ages
are more metal-poor and exhibit vertical gradients in excellent agreement with the results 
derived from thick disk stars in previous studies. More importantly, it seems that stars 
of the oldest ages have radial metallicity gradients comparable to those yielded by stars 
of all ages at large disk heights, where the thick disk stars are expected to dominate. 
In a parallel work, we show that the mean age of stars increase with disk heights, 
and stars of the oldest ages dominate the populations at large heights (Xiang et al. in prep.). 
This suggests that disk stars of the oldest ages are in fact thick disk stars. We thus propose 
that age is a natural, clear criterion that distinguishes the thin and thick disk stars: 
stars of the oldest ages belong to the thick disk while those of younger ages are from 
the thin disk.  

The above discussion suggests that the thick disk corresponds an earlier 
phase of the disk formation via processes that seem to be quite different from 
those responsible for the formation of the younger, thin disk. 
This is in conflict with the recent suggestion of Bovy et al. (2012), who suggest that 
the thin and thick disks are probably not two distinct components of the disk in terms of origin 
based on their finding that in the [Fe/H] -- [$\alpha$/Fe] plane, stellar populations 
of the largest scale heights have the smallest scale lengths. 
On the other hand,  Haywood et al. (2013) have recently studied the age -- [Fe/H] 
and age -- [$\alpha$/Fe] relations of 1111 FGK stars in the solar neighborhood. 
They find that thick disk stars show a tight correlation of [$\alpha$/Fe] and [Fe/H] with age. 
They thus suggest that thick disk stars have an origin different to that 
of the younger, thin disk stars. Haywood et al. (2013) suggest that the older, thick 
disk formed from well-mixed interstellar medium over a time scale of 4 -- 5\,Gyr at 8\,Gyr ago, 
the $inner$ thin disk formed later, with the initial chemical conditions set by the 
youngest thick disk. While the $outer$ thin disk is different, as it has began to form at 10\,Gyr ago. 
Our results suggest the idea that the gas is probably well-mixed in the radial direction $11$\,Gyr 
ago, but probably not so in the vertical direction. Stars of age between 8 and 11\,Gyr 
begin to show some radial gradients, and they probably a mixture of thin and  
thick disk stars. Given the relatively large uncertainties of our age estimates, the exact 
epoch when the thick disk formation ceased and the thin disk formation prevailed remains 
to be determined. The epoch should be sometime around 11\,Gyr but in any case 
earlier than 8\,Gyr. The transition must be rapid since we see an abrupt change 
in the slopes of both radial and vertical metallicity gradients between stars of ages 
older than 11\,Gyr and those of ages between 8 and 11\,Gyr. Thus, we suspect that 
thick disk may have formed slightly quicker than claimed by Haywood et al. 
With regard to the thin disk, there is no strong evidence to group them into those from the $inner$ 
and those from the $outer$ parts of the thin disk in discussing their possible formation scenarios. 

\subsection{Constraining the disk formation scenarios}
With regard to the thick disk formation, as discussed above, our results prefer a 
violent, fast formation process. 
One possibility is the scenario of a fast, highly-turbulent gas-rich 
merger formation as proposed by Brook et al. (2004, 2005). The 
scenario seems to be preferred by a number of authors in explaining their results 
of metallicity gradients (e.g. Cheng et al. 2012b; Schlesinger et al. 2012, 2014; 
Kordopatis et al. 2011; Boche et al. 2013; Hayden et al. 2014). 
Nevertheless, we note that based on their simulations, Brook et al. suggest that 
the thick disk formation occurred so fast $\sim$12\,Gyr ago that there should be 
no metallicity gradients at all in both the radial and vertical directions. However, 
On the other hand, significant vertical gradients for thick disk stars have been 
detected by many studies, including the current one.  
This may imply that the thick disk formation process of Brook et al. is probably  
too fast and violent. In fact, Gilmore et al. (1989) point out 
that a slow, pressure-supported gas collapse following the formation 
of extreme Population II system could form a thick disk. 
Based on the hydrodynamical models of Larson (1976), 
disk formed in such a way shows weak radial metallicity gradients 
in the solar neighborhood but strong vertical gradients because 
the vertical flow dominates when the gas settles down to disk at that epoch.  
Such a slow, pressure-supported collapse formation scenario 
seems to be consistent with our results, though better, quantitive 
models/simulations are needed to demonstrate this. 

It has also been proposed that the thick disk is produced by stellar radial migration 
(e.g. Ro$\breve{s}$kar et al. 2008;  Sch\"onrich \& Binney 2009b; Loebman et al. 2011). 
Our results show that at $|Z| < $1.0\,kpc, the radial gradients of the old, 
thick disk stars, as well as their trend of variations with $|Z|$, are significantly different 
from those derived from stars of age bins 8 -- 11\,Gyr and 6 -- 8\,Gyr. In addition, 
for a given $|Z|$, there is a turning in the temporal variations of radial gradients, 
at an epoch of $\sim$8\,Gyr, after which the gradients become flatter as age decreases. 
These results are difficult to explain by radial migration since radial migration 
is expected to reduce the radial gradients (e.g. Loebman et al. 2011; Kubryk et al. 2013), 
and as such one expects a continuous rather than abrupt temporal variations.  
Note that this does not imply there is no radial migration, but that the radial 
migration is probably not the dominant mechanism that shape the properties of metallicity 
gradients of concern. The observed metallicity gradients are likely the results of 
both gas infall/inflow, which steepens the gradients, and stellar migration, 
which flattens the gradients. Based on results presented in the current work, 
it seems to us that gas infall/inflow probably dominates in shaping the metallicity gradients of the disk 
and their temporal variations. In fact, there is simulations suggesting that stellar radial
migration does not change the trend of variations of radial metallicity gradients (Curir et al. 2014). 

For the thin disk evolution, it has been suggested that stellar radial migration has played 
an important role considering that it successfully explain the age -- metallicity 
distribution and metallicity -- velocity relations of stars in the solar 
neighborhood (e.g. Sch\"onrich \& Binney 2009b; Lee et al. 2011; Loebman et al. 2011). 
Our current analyses show that the radial gradients deduced from young, thin disk 
stars flatten with increasing $|Z|$. Similarly, the vertical gradients flatten as $R$ increases. 
These trends are probably the consequence of outward stellar radial migration,  
since as a star migrates outward, its altitude is expected to increase significantly (e.g. Minchev et al. 2012). 
In addition, both metal-rich stars of very old ages (e.g. 8 -- 11\,Gyr), and metal-poor 
stars of very young ages (e.g. $<4$\,Gyr) are present in the solar neighborhood (Fig.\,14). 
A direct examination of the age -- metallicity distribution shows similar results. 
This is clear signature of stellar radial migration, which brings inner,  
metal-rich stars outward, and outer, metal-poor stars inward.  In the panels of Fig.\,14 
showing the results of stars of the oldest ages, there are very few old metal-rich stars 
compared to panels showing the results of stars of younger age bins. 
This again suggests that the thick disk does not result from radial migration of thin disk stars. 
It seems that 8\,Gyr is a specific epoch in the formation history of the thin disk, since 
it is around this epoch that both the radial and vertical metallicity gradients reach their 
maximum (negative) values. We suspect that around 8\,Gyr, the gas accretion/inflow 
rate in the solar neighbourhood reached the maximum.

\section{Summary}
In this work, we have defined a sample of 297\,042 MSTO stars of effective temperature 
range of $5400 < T_{\rm eff} < 7500$\,K from the LSS-GAC survey.
Most of the sample objects  are disk stars of metallicity ${\rm [Fe/H]} > -1.0$\,dex. 
The sample is used to study the metallicity gradients of the Galactic disk in the anti-center direction.
For any age between 2 and 13\,Gyr, our sample encompasses stars 
spreading over almost the whole metallicity range of typical disk stars. 
We estimate distances and ages of the sample stars by isochrone fitting, 
and achieve an accuracy of about 20 per cent in distance and 30 per cent in age. 
The sample covers a contiguous space volume from 7.5 to13.5\,kpc in 
Galactocentric radial distance $R$ and from $-2.5$ to 2.5\,kpc in disk height $Z$.  

To correct for the sample selection effects, we first apply a CMD weight to  
the individual sample star, leads us to a magnitude limited sample. 
We then characterize the potential biases of metallicity gradients deduced from a 
magnitude limited sample utilizing mock data generated with Monte Carlo simulations 
of the Galactic disk. We find that the radial metallicity gradients derived from a sample of  
stars in a narrow range of ages reproduce the underlying true values well. 
However, if the sample includes stars over a wide range of ages, then the 
radial gradients derived will be severely biased to larger values (i.e. less steep for 
negative radial gradients which is the case for the real Galactic disk). 

Both the radial and the vertical metallicity gradients are found to 
show significant spatial and temporal variations.The radial gradients, 
$\Delta$[Fe/H]/$\Delta$$R$,  deduced from stars of the oldest ages ($> 11$\,Gyr) 
are essentially $zero$ at all heights above the Galactic disk plane, 
while those deduced from stars in younger age bins are always negative at all heights. 
The vertical gradients, $\Delta$[Fe/H]/$\Delta|Z|$, 
deduced from stars of the oldest ages are about $-0.1$\,dex\,kpc$^{-1}$, 
with some marginal variations with $R$, while those deduced from stars in younger 
age bins show large variations with $R$. Based on the results, we infer that the 
Galactic disk formation may have experienced at least two distinct phases. In the earlier phase 
($\sim$11\,Gyr ago), gas settled down to the disk mainly along the direction 
perpendicular to and produced stars with a negative vertical gradient. 
The gas is well-mixed in the radial direction such that the (thick disk) stars 
formed exhibit no radial gradients. In the later phase, the disk build up mainly 
by gas accretion/inflow in the disk, yielding negative radial gradients. 
Our metallicity gradients deduced from stars of the oldest ages are well consistent with 
the previous determinations using samples of thick disk stars selected via 
either spatial or chemical criteria (e.g. Chen et al. 2011; Carrell et al. 2012; Cheng et al. 2012b). 
We thus suggest that the thick component of the Galactic disk corresponds to an earlier phase 
of disk formation, while the thin component corresponds a 
later phase of disk buildup. We propose that stellar age serve as a natural, physical 
quantity to distinguish the thin and thick disk stars.   
Given that the thick disk shows negative vertical gradients, the gas-rich 
merger scenario of Brook et al. (2004, 2005) is not preferred, since in this scenario 
the thick disk forms so fast that it does not produce significant negative vertical gradient. 
On the other hand, a slow, pressure-supported collapse of gas 
following the formation of the Population II system, as first outlined 
by Gilmore et al. (1989), is more consistent with the current observations. 
There is an abrupt, significant change in both the 
radial and vertical metallicity gradients from the thick to the thin disk, 
suggesting that stellar radial migration (e.g. Sch\"onrich \& Binney, 2009b; Loebman et al. 2011) 
might not play an important role in the thick disk formation. 

At the earliest epochs, the radial gradients steepen with decreasing age, 
reaching maxima around 7 -- 8\,Gyr ago, after which the gradients flatten. 
Such trends of temporal variations are seen at all disk heights below 1.5\,kpc. 
Similar trends of variations are also observed in the vertical gradients. 
It seems that 8\,Gyr is a specific epoch in the buildup of the thin disk. 
It is possible that around this epoch, the gas accretion/inflow rate in the solar 
neighborhood reaches a maximum. This does not necessarily imply that the 
thin disk started to form only 8\,Gyr ago. There is likely a transition period, 
probably before 8\,Gyr ago, during which the disk buildup mode 
(in term of gas infall, accretion/inflow) changed from that of thick disk to 
the one of thin disk. Due to the relatively large uncertainties of our current 
age estimates, the exact epoch that this transition occurred remains to be determined. 
The fact that the radial metallicity gradients of the thin disk steepen with decreasing age
implies that stellar radial migration is also unlikely to have played a decisive role 
in the formation and evolution of metallicity gradients of the thin disk. 

A two-phase formation of the Galactic disk has been previously suggested by 
Haywood et al. (2013). The observations presented here however provide more 
details to this picture. For example, the presence of negative vertical metallicity 
gradients at the earliest epochs seem to suggest that gas infall at those earliest 
epochs are mainly along the direction perpendicular to the disk. The results also 
point to a slow, pressure-supported collapse of gas following the formation of the 
Population II system (Gilmore et al. 1989) for the thick disk formation. 
Our data also seem to suggest that the formation of thin disk started earlier than 
proposed by Haywood et al., probably sometime between 8 and 11\,Gyr.
Finally, Haywood et al. suggest that the inner and outer disks may have different 
origins, with the outer disk started to form earlier than the inner disk. 
They suggest that stars in the inner (metal-rich) thin disk are all younger than 8\,Gyr, while in the outer 
(metal-poor) thin disk there are stars as old as 9 -- 10\,Gyr. However, our analysis 
seem to suggest that dividing the thin disk into the inner and outer 
parts in terms of different formation epochs is probably not necessary and substantiated by the data.

\vspace{7mm} \noindent {\bf Acknowledgments}{
This work is supported by National Key Basic Research
Program of China 2014CB845700.
Guoshoujing Telescope (the Large Sky Area Multi-Object Fiber
Spectroscopic Telescope LAMOST) is a National Major Scientific
Project built by the Chinese Academy of Sciences. Funding for
the project has been provided by the National Development and
Reform Commission. LAMOST is operated and managed by the National
Astronomical Observatories, Chinese Academy of Sciences. 
This work is also supported by National Natural Science Foundation of China
(Grant No. 11473001). M.S.X. thanks Professor Jian-Ning Fu for  
providing the LAMOST data of Kepler fields. B.Q.C acknowledges partial funding
from China Postdoctoral Science Foundation 2014M560843.}

\label{lastpage}

\end{document}